\documentclass[12pt]{article}

\usepackage{verbatim}
\usepackage{amsmath}
\usepackage{amssymb}
\usepackage{graphicx}
\usepackage{cite}
\usepackage{subfig}
\usepackage{setspace}
\usepackage{url}
\usepackage[percent]{overpic}
\usepackage{slashed}
\usepackage{authblk}

\usepackage{xspace}
\newcommand{\Fermi}[0]{\textit{Fermi}\xspace}

\newcommand{\sigv}[0]{\ensuremath{\langle\sigma v\rangle}\xspace}
\newcommand{\sigvm}[0]{\ensuremath{\langle\sigma v\rangle_{\rm{UL}}}\xspace}

\newcommand{\unit}[1]{\ensuremath{\mathrm{\,#1}}\xspace}

\newcommand{\GeV}{\unit{GeV}}
\newcommand{\MeV}{\unit{MeV}}
\newcommand{\degree}{\unit{^{\circ}}}

\usepackage{fullpage}
\usepackage{hyperref}
\def\tev{\,{\rm TeV}}
\def\gev{\,{\rm GeV}}
\def\ie{{\it i.e.}}
\def\eg{{\it e.g.}}

\def\to{\rightarrow}
\def\Fermi{\,{\it Fermi}}

\title{Complementarity of Dark Matter Searches in the pMSSM}
\date{}
\author[1]{M. Cahill-Rowley}
\author[2]{R. Cotta}
\author[3]{A. Drlica-Wagner}
\author[1]{S. Funk}
\author[1]{J.L. Hewett}
\author[4,5]{A. Ismail}
\author[1]{T.G. Rizzo}
\author[1]{M. Wood}
\affil[1]{SLAC National Accelerator Laboratory, Menlo Park, CA, USA\footnote{mrowley, funk, hewett, rizzo, mdwood@slac.stanford.edu}}
\affil[2]{University of California, Irvine, CA, USA\footnote{cottar@uci.edu}}
\affil[3]{Fermi National Accelerator Laboratory, Batavia, IL, USA\footnote{kadrlica@fnal.gov}}
\affil[4]{Argonne National Laboratory, Argonne, IL, USA\footnote{aismail@anl.gov}}
\affil[5]{University of Illinois, Chicago, IL, USA}

\begin{document}

\rightline{\vbox{\halign{&#\hfil\cr
&SLAC-PUB-15862\cr
}}}


{\let\newpage\relax\maketitle}

\begin{abstract}
As is well known, the search for and eventual identification of dark matter in supersymmetry requires a simultaneous, multi-pronged approach with important roles played by the LHC 
as well as both direct and indirect dark matter detection experiments. We examine the capabilities of these approaches in the 19-parameter p(henomenological)MSSM which provides a general framework for complementarity studies of neutralino dark matter.
We summarize the sensitivity of dark matter searches at the 7, 8 (and eventually 14) TeV LHC, combined with 
those by \Fermi, CTA, IceCube/DeepCore, COUPP, LZ and XENON. The strengths and weaknesses of each of these techniques are examined and contrasted  
and their interdependent roles in covering the model parameter space are discussed in detail. We find that these approaches explore orthogonal territory and that advances in each are necessary to cover the Supersymmetric WIMP parameter space.  We also find that different
experiments have widely varying sensitivities to the various dark matter annihilation mechanisms, some of which would be completely excluded by null results from these experiments.
\end{abstract}

\section{Introduction and Overview of the pMSSM}

Determining the identity of dark matter (DM) is one of the most pressing issues before us today.  Multiple observations reveal that roughly 85\% of the matter in the universe is electromagnetically inert, and the local density of dark matter is known to within a factor of two. Cosmological considerations reveal a handful of its properties, yet it may take many forms, and numerous theories hypothesize dark matter particles of various types.  A promising class of dark matter candidates is Weakly Interacting Massive Particles (WIMPs) which could have thermally frozen out in the early universe in a manner that yields the relic density observed by experiment today.  WIMPs naturally appear in many extensions of the Standard Model (SM) that resolve the gauge hierarchy, with the most notable example being supersymmetry (SUSY).  

Several mechanisms allow for the search for WIMP dark matter: ($i$) direct detection where WIMPs elastically scatter off nuclei, ($ii$) indirect detection where WIMPs annihilate into a pair of SM particles, and ($iii$) the direct production of WIMPs in high energy colliders.  In this paper we investigate the complementary roles these three search techniques play in the quest to discover dark matter.  We employ the phenomenological Minimal Supersymmetric Standard Model (pMSSM) \cite{Djouadi:1998di,Berger:2008cq} as our tool to examine the coverage of WIMP parameter space by each technique.  We find that the methods explore orthogonal territory and that advances in all three techniques are necessary to cover the supersymmetric WIMP sector.  Here, we first outline the salient features of the pMSSM, examine each WIMP detection experiment in turn, and then draw conclusions from the combined results.

One of the main reasons that R-parity conserving supersymmetry is attractive is the prediction that the lightest SUSY particle (LSP) is stable and may be 
identified as a thermal dark matter candidate if it is both electrically neutral and a color singlet. Frequently in the MSSM, the LSP is associated with the lightest neutralino, $\chi_1^0$. 
While DM searches are directly focused on the nature of the LSP itself, 
the properties of the full spectrum of superparticles, and of the extended SUSY Higgs sector, also play important roles. Thus it is inappropriate to 
completely separate DM searches from the exploration and examination of the rest of the SUSY spectrum. However, even in the simplest SUSY scenario, the MSSM, the 
number of free parameters ($\sim$100) is simply too large to perform a study in all generality; we thus need to restrict our view without losing any 
relevant physics. One approach is to assume the existence of a high-scale theory with a handful of parameters (such as mSUGRA{\cite {SUSYrefs}) from which 
all the properties of the TeV scale sparticles can be determined and studied. While this method is valuable and predictive, these scenarios 
are somewhat phenomenologically limiting and are under ever-increasing tension with a wide range of experimental data including, in some cases, the $\sim 126$ GeV 
mass of the recently discovered Higgs boson{\cite {ATLASH,CMSH}}.   

One way of circumventing these limitations is to examine the more general 19-parameter pMSSM{\cite{Djouadi:1998di,Berger:2008cq}}.  The pMSSM is the most general version 
of the R-parity conserving MSSM that satisfies several data-driven constraints: ($i$) no new phase appearing in the soft-breaking parameters, \ie, CP 
conservation, ($ii$) Minimal Flavor Violation at the electroweak scale such that the CKM matrix drives flavor mixing, ($iii$) degenerate first and second generation 
soft sfermion masses, ($iv$) negligible Yukawa couplings and associated A-terms for the first two generations. In particular, note that the pMSSM contains no theoretical 
assumptions about physics above the TeV scale, {\it e.g.,} the nature of SUSY breaking or grand unification.  This allows for the capture of electroweak scale phenomenology for which a UV-complete theory may not yet exist. When we impose the constraints ($i$)-($iv$), the number of free parameters in the MSSM at the TeV-scale decreases from 105 to 19 for the case 
of a neutralino LSP (or 20 including the gravitino mass as an additional parameter when it plays the role of the LSP{\footnote {In this work we will 
limit our discussion to the case of neutralino LSPs.}}).  

To study the pMSSM, we essentially throw darts into this large space, generating many millions of random model points (using SOFTSUSY{\cite{Allanach:2001kg}} and  
checking for consistency using SuSpect{\cite{Djouadi:2002ze}}), with each point corresponding to a specific set of values for the parameters.
The ranges of the 19 pMSSM parameters employed in the analysis below are presented in Table~\ref{ScanRanges}. The lower and upper bounds used for the ranges in our scan
were chosen to be essentially consistent with Tevatron and LEP data and to have kinematically-accessible sparticles at the 14 TeV high-luminosity LHC, respectively. 
We do not assume that the thermal relic density as calculated 
for the neutralino LSP necessarily saturates the WMAP/Planck value{\cite{Komatsu:2010fb}} in order to allow for the possibility of multi-component DM. 
For example, axions introduced to solve the strong CP problem might make up a substantial amount of DM. 
Decay patterns of the SUSY partners and the extended Higgs sector are calculated using a privately modified version of 
SUSY-HIT{\cite{Djouadi:2006bz}}. 
Each individual model is then subjected to a large set of collider, flavor, precision measurement, dark matter and theoretical constraints~\cite{us1}.  
Roughly 225k models with a neutralino LSP survive this initial selection and can then be used for further physics studies.  We note that this model set was
generated before the discovery of the Higgs boson and that approximately 20\% of the sample predicts the correct Higgs mass.  We will discuss possible effects
of this below.
We have recently performed a detailed study of the signatures for the pMSSM at the 7 and 8 (and eventually 14) TeV 
LHC {\cite {us1,us2,Cahill-Rowley:2013yla}} and include these results here.

\begin{table}
\centering
\begin{tabular}{|c|c|} \hline\hline
$m_{\tilde L(e)_{1,2,3}}$ & $100 \gev - 4 \tev$ \\ 
$m_{\tilde Q(q)_{1,2}}$ & $400 \gev - 4 \tev$ \\ 
$m_{\tilde Q(q)_{3}}$ &  $200 \gev - 4 \tev$ \\
$|M_1|$ & $50 \gev - 4 \tev$ \\
$|M_2|$ & $100 \gev - 4 \tev$ \\
$|\mu|$ & $100 \gev - 4 \tev$ \\ 
$M_3$ & $400 \gev - 4 \tev$ \\ 
$|A_{t,b,\tau}|$ & $0 \gev - 4 \tev$ \\ 
$M_A$ & $100 \gev - 4 \tev$ \\ 
$\tan \beta$ & 1 - 60 \\
\hline\hline
\end{tabular}
\caption{Scan ranges for the 19 
parameters of the pMSSM with a neutralino 
LSP employed in the present analysis. 
The parameters are scanned with flat priors, and we expect this choice to have little qualitative impact on our results~\cite{us}.}
\label{ScanRanges}
\end{table}

As a result of our scan ranges for the electroweak gauginos (chosen for compatibility with LEP data and to enable phenomenological studies at the 14 TeV LHC), the 
LSPs in our model sample are typically very close to being in a pure electroweak eigenstate as the off-diagonal elements of the chargino and neutralino mass matrices are 
at most $\sim M_W$. Figure~\ref{fig0} presents some properties of the nearly pure eigenstate LSPs (defined here as a single electroweak eigenstate comprising over 90\% of 
the mass eigenstate). The left panel displays the distribution of the LSP mass for nearly pure bino, wino, and Higgsino LSPs, while the right-hand panel shows the 
corresponding distribution for the predicted LSP thermal relic density. Note that the LSP masses lie below $\sim 2$ TeV in all models; this is due to our choice of scan ranges as the entire SUSY spectrum must be lighter than $\sim 4$ TeV and heavier than the LSP (by definition), and this becomes increasingly improbable with increasing LSP mass. In addition, the relic density upper limit becomes increasingly difficult to satisfy at larger LSP masses. Similarly, due to LEP and relic density constraints, 
none of our models have LSP masses below $\sim$40 GeV. The fraction of models where the LSP is nearly a pure bino eigenstate 
is found to be rather low in this model sample since such models 
generally lead to too high a value for the relic density unless they co-annihilate with another sparticle, happen to be close to a ($Z,h,A$) funnel region, or have 
a suitable Higgsino admixture. Note that only in the rightmost bin of the right panel is the relic density approximately saturating the WMAP/Planck thermal relic value. 
These LSP properties will be of particular importance in the discussions that follows.

\begin{figure}[htbp]
\centerline{\includegraphics[width=3.5in]{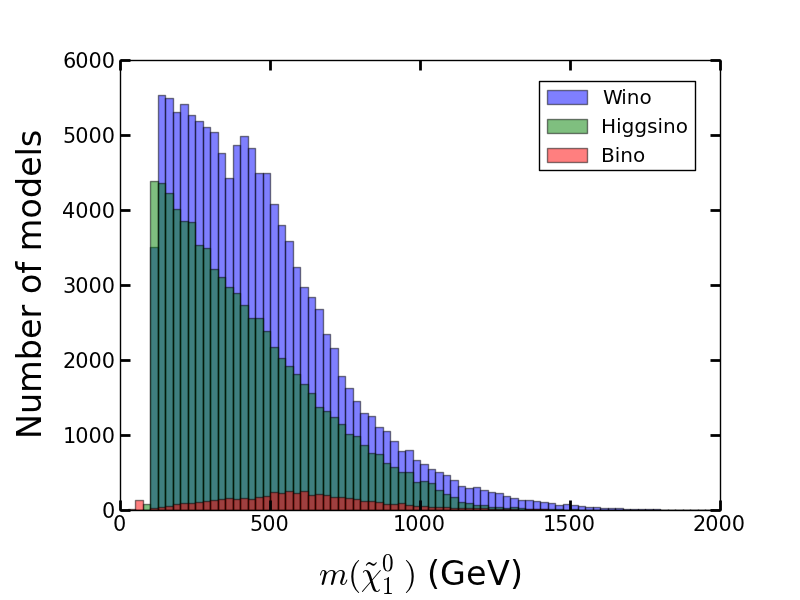}
\hspace{-0.50cm}
\includegraphics[width=3.5in]{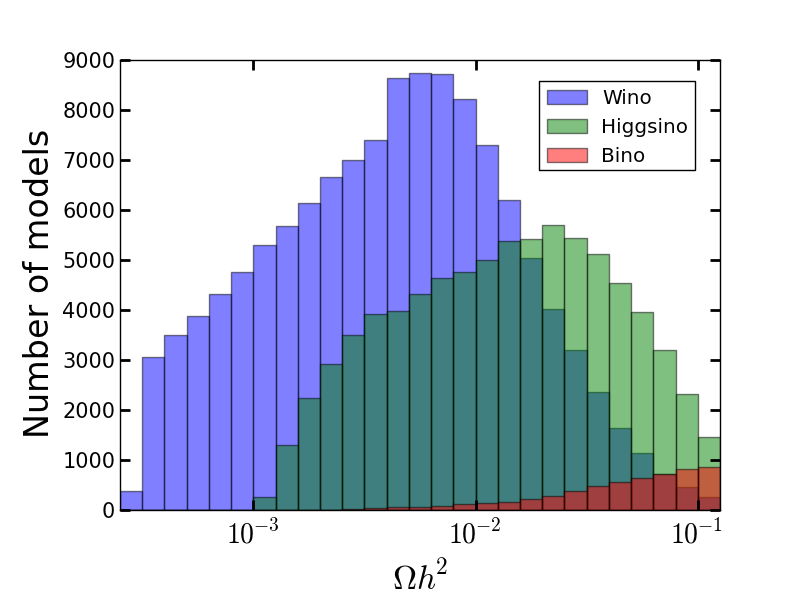}}
\vspace*{-0.10cm}
\caption{Distribution of the LSP masses (left) and predicted relic density (right) for the neutralino LSPs that are almost pure weak eigenstates in our model sample.}
\label{fig0}
\end{figure}

Figure~\ref{fig00} shows the thermal relic density generated by the LSP in our pMSSM models as a function of the LSP mass with the color-coding reflecting their electroweak 
eigenstate content. There are many items to note here that will be important for later consideration. Essentially every possible known mechanism to obtain (or lie below) the 
WMAP/Planck relic density is present: ($i$) The set of models with low LSP masses (forming `columns' on the left-hand side of the figure) correspond to bino-Higgsino admixtures which achieve a sufficiently low relic density by resonant annihilation through the $Z,h$-funnels; these sometimes are pure binos if the Higgsino fraction is very small{\footnote {Here again, `pure' means having an eigenstate fraction $\geq 90\%$. Points shown as bino-wino, bino-Higgsino, or wino-Higgsino mixtures have 
less than $2\%$ Higgsino, wino, or bino fraction, respectively. Mixed points have no more than $90\%$ and no less than $2\%$ of each component.}}. ($ii$) The bino-Higgsino LSPs saturating the relic density in the upper left region of the figure are of the so-called `well-tempered' variety. 
($iii$) the pure bino models in the upper middle region of the Figure are bino co-annihilators (mostly with sleptons) or annihilate resonantly through the $A$-funnel. ($iv$) The green (blue) bands are pure Higgsino (wino) models 
that saturate the relic density bound (using perturbative calculations which do not include the Sommerfeld enhancement effect{\footnote {The Sommerfeld enhancement can significantly deplete the relic density of wino LSPs heavier than $\sim$ 1 TeV, while Higgsino and light wino LSPs are relatively unaffected~\cite{Hryczuk:2010zi}. Bino LSPs do not exhibit the effect because they can't exchange gauge bosons. Including the enhancement would increase the low-velocity annihilation cross section for heavy winos, lowering their predicted relic density but increasing their present-day annihilation cross section. Since the average velocity today is lower than during freeze-out, we would naively expect that including the enhancement would strengthen the limits on heavy wino LSPs. We will see that CTA is already able to exclude models with heavy winos in our perturbative calculation; we therefore expect that including the enhancement would minimally affect our conclusions.}}) near $\sim 1(1.7)$ TeV and have very low relic densities for lighter LSP masses. Wino-Higgsino hybrids are seen to lie between these two cases as expected. ($v$) A smattering of models with additional (or possibly multiple) annihilation channels are loosely distributed in the lower right-hand corner of the Figure. As we will see, many of the searches for DM are particularly sensitive to one or more of these LSP categories.

\begin{figure}[htbp]
\centerline{\includegraphics[width=7.0in]{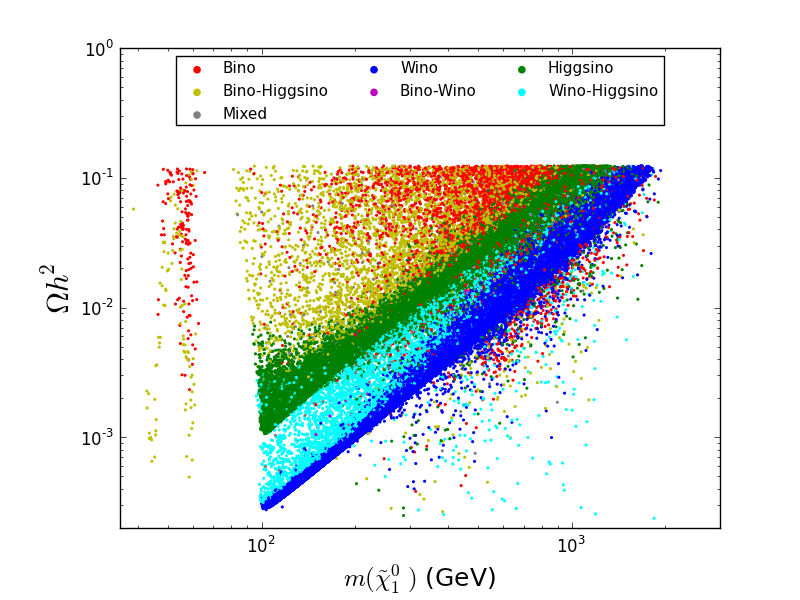}}
\vspace*{-0.10cm}
\caption{Thermal relic density generated in our pMSSM model set as a function of the LSP mass, color-coded by the electroweak properties of the  
LSP as indicated and discussed in the text.}
\label{fig00}
\end{figure}

\section{LHC Searches}

We begin with a short overview of the searches for the pMSSM at the 7 and 8 TeV LHC~\cite{us1,us2,Cahill-Rowley:2013yla}. In general, our approach is to closely follow the 
suite of ATLAS SUSY analyses but also to supplement these with several searches performed by CMS; the analyses included in our study are briefly summarized in Table~\ref{SearchList}. In addition, we include the searches for heavy neutral SUSY Higgs decaying to $\tau^+\tau^-$ by CMS~\cite{Chatrchyan:2012vp} and measurements of the rare decay mode  
$B_s\to \mu^+\mu^-$ discovered by CMS and LHCb~\cite{BSMUMU}. Both of these additional searches play distinct but important roles in restricting the pMSSM parameter space. 
We have implemented every relevant ATLAS SUSY search publicly available as of the beginning of March 2013 and also the more recent 20 fb$^{-1}$ 2-6 jets + MET 
analysis. The LHC results for our model sets (including the neutralino LSP models considered in this paper) appear in detail in our companion 
papers on both neutralino and gravitino LSP SUSY searches~\cite{Cahill-Rowley:2013yla, newpaper}.

A brief summary of our procedure is as follows: We generate SUSY events for each model using PYTHIA 6.4.26~\cite{Sjostrand:2006za} and PGS 4~\cite{PGS}, which we have 
modified to, e.g., correctly deal with gravitinos, hadronization of stable colored sparticles, multi-body decays and ATLAS b-tagging methods. We then scale our event rates to 
NLO by calculating the relevant K-factors with Prospino 2.1~\cite{Beenakker:1996ch}. The individual searches are then implemented using a custom analysis 
code{\cite {us}}, following the published cuts and selection criteria of ATLAS as closely as possible. Our code is validated for each of the many search regions 
in every analysis employing the benchmark model points provided by ATLAS (and CMS). Models are then excluded using the $95\%$ $CL_s$ limits as employed by ATLAS. Note that 
these analyses are performed {\it without} imposing the Higgs mass constraint, $m_h=126 \pm 3$ GeV (combined experimental and theoretical errors) so that we can understand 
its impact on the search results. Roughly $20\%$ of models in the neutralino model set (before the LHC SUSY searches are applied) predict a Higgs mass in the above 
range. While there is some variation amongst the individual searches we find that, once combined, the total fraction of our models surviving (or excluded by) the set of all 
LHC searches is to an excellent approximation {\it independent} of whether or not the Higgs mass constraint has been applied. Conversely, the $\sim 20\%$ fraction of the 
neutralino models predicting the correct Higgs mass is also found to be approximately independent of whether or not the SUSY searches have been applied. This combined result 
is very powerful, demonstrating the current approximate decoupling of SUSY search results from the discovery of the Higgs boson and this allows us to employ the  
entire model set for SUSY studies with some validity. After all of these searches are applied, we find that 45.5\% of the pMSSM model sample is excluded, leaving 54.5\% of the set as viable models.

\begin{table}
\centering
\begin{tabular}{|l|l|l|} \hline\hline
Search  & Energy &   Reference     \\
\hline
2-6 jets & 7 TeV & ATLAS-CONF-2012-033   \\
multijets & 7 TeV & ATLAS-CONF-2012-037    \\
1 lepton  & 7 TeV & ATLAS-CONF-2012-041    \\

2-6 jets   & 8 TeV &   ATLAS-CONF-2012-109   \\
multijets   & 8 TeV &  ATLAS-CONF-2012-103    \\
1 lepton     & 8 TeV &  ATLAS-CONF-2012-104   \\
SS dileptons & 8 TeV &  ATLAS-CONF-2012-105    \\
2-6 jets & 8 TeV   & ATLAS-CONF-2013-047  \\

Gluino $\to$ Stop/Sbottom   & 7 TeV & 1207.4686  \\
Very Light Stop  & 7 TeV & ATLAS-CONF-2012-059   \\
Medium Stop  & 7 TeV & ATLAS-CONF-2012-071 \\
Heavy Stop (0l)  & 7 TeV & 1208.1447 \\
Heavy Stop (1l)   & 7 TeV & 1208.2590  \\
GMSB Direct Stop  & 7 TeV & 1204.6736   \\
Direct Sbottom & 7 TeV & ATLAS-CONF-2012-106  \\
3 leptons & 7 TeV & ATLAS-CONF-2012-108  \\
1-2 leptons & 7 TeV & 1208.4688  \\
Direct slepton/gaugino (2l)  & 7 TeV & 1208.2884  \\
Direct gaugino (3l) & 7 TeV & 1208.3144  \\
4 leptons & 7 TeV & 1210.4457  \\
1 lepton + many jets & 7 TeV & ATLAS-CONF-2012-140  \\
1 lepton + $\gamma$ & 7 TeV & ATLAS-CONF-2012-144  \\
$\gamma$ + b & 7 TeV & 1211.1167  \\
$\gamma \gamma $ + MET ($\tilde{G}$ LSP) & 7 TeV & 1209.0753  \\

Medium Stop (2l) & 8 TeV & ATLAS-CONF-2012-167   \\
Medium/Heavy Stop (1l) & 8 TeV & ATLAS-CONF-2012-166  \\
Direct Sbottom (2b) & 8 TeV & ATLAS-CONF-2012-165  \\
3rd Generation Squarks (3b) & 8 TeV & ATLAS-CONF-2012-145  \\
3rd Generation Squarks (3l) & 8 TeV & ATLAS-CONF-2012-151  \\
3 leptons & 8 TeV & ATLAS-CONF-2012-154  \\
4 leptons & 8 TeV & ATLAS-CONF-2012-153  \\
Z + jets + MET & 8 TeV & ATLAS-CONF-2012-152  \\

HSCP      & 7 TeV  &  1205.0272    \\
Disappearing tracks  & 7 TeV  &  ATLAS-CONF-2012-111  \\
Muon + Displaced Vertex ($\tilde{G}$ LSP) & 7 TeV  & 1210.7451  \\
Displaced Dilepton ($\tilde{G}$ LSP) & 7 TeV  & 1211.2472  \\

$B_s \to \mu \mu$ & 7 + 8 TeV  & 1211.2674  \\
$A/H \to \tau \tau$ & 7 + 8 TeV  & CMS-PAS-HIG-12-050  \\

\hline\hline
\end{tabular}
\caption{Simulated LHC SUSY searches that have been applied to our pMSSM model set. $54.5\%$ of the models with a neutralino 
LSP survive all of these searches and remain viable. This is found to be approximately independent of the Higgs mass constraint.}
\label{SearchList}
\end{table}

Figure~\ref{fig1} demonstrates the power obtained by combining the set of LHC analyses to constrain the pMSSM neutralino LSP models. In 
particular, this figure shows the fraction of models having a given sparticle and LSP mass that are excluded by the combined LHC searches. Since the values of the 
masses of the squarks and gluinos, the lightest stops and sbottoms, and the LSP itself are of particular interest, we concentrate on these specific  
quantities in this figure. In the upper left-hand panel the coverage of the gluino-LSP mass plane by the LHC searches is displayed; the white line represents the $95\%$ CL search limit on a simplified model with a gluino NLSP, neutralino LSP, and all other sparticles decoupled, as obtained by ATLAS from their 20 fb$^{-1}$ 2-6 jets plus MET analysis~\cite{TheATLAScollaboration:2013fha}. We see that this is very roughly the same as the black region excluded in the pMSSM. Note however that the pMSSM exclusion is slightly stronger than the simplified model limit for lighter gluino masses, while being somewhat weaker than the simplified model limit in the heavy gluino region. Of course, the fact that the other sparticles are generically not decoupled in the pMSSM means that the gluinos exhibit many different decay patterns, some of which are rather insensitive to the jets + MET search. It is then interesting that the pMSSM exclusion resulting from the combination of multiple searches is similar to the limit from the jets + MET search in the simplified model scenario. Generally, models with heavy gluinos that are not excluded, despite being below the simplified model limit, have decays through stops, both on-shell and off-shell. The upper right-hand panel shows the corresponding coverage in the gluino-lightest squark mass plane with a simplified model line, again from the $\sim 20$ fb$^{-1}$ 2-6 jets plus MET analysis~\cite{TheATLAScollaboration:2013fha}. In this case, the simplified model assumes that the LSP is massless and that the 8 squarks of the first two generations are degenerate, neither of which are common occurances in the pMSSM. As a result, it is no surprise that our excluded region is not well described by the simplified model. While most models with 
rather light squarks and/or gluinos are observed to be excluded by the combined LHC searches, it is clear that models with squarks and/or gluinos below $\sim 700-750$ GeV 
still remain viable.

The lower two panels display the lightest stop/sbottom - LSP mass planes, again including the corresponding ATLAS simplified model limits~\cite{ATLAS3glimits}. 
Here we see several things, the most important being that the region of coverage in these two planes in the pMSSM differs substantially from either simplified model limit. As we describe in~\cite{Cahill-Rowley:2013yla}, this is because stops and sbottoms can typically decay to either a neutralino or a chargino (since the LSP is most commonly a wino or Higgsino multiplet), producing a mixture of final states. In particular, the LHC sensitivity to stops improves substantially when the branching fraction for $\tilde{t}_1 \to b \chi_1^+$ is appreciable, since the resulting hard b-jets are relatively easy to distinguish from the $t\bar{t}$ background. This effect is most striking in the compressed spectrum region where the stop simplified model limit is far weaker than the pMSSM exclusion. In the highly-compressed region, the exclusion reach results mainly from generic jets + MET searches, as the b-jet pT becomes too low for a reasonable tagging efficiency.

\begin{figure}[htbp]
\centerline{\includegraphics[width=3.5in]{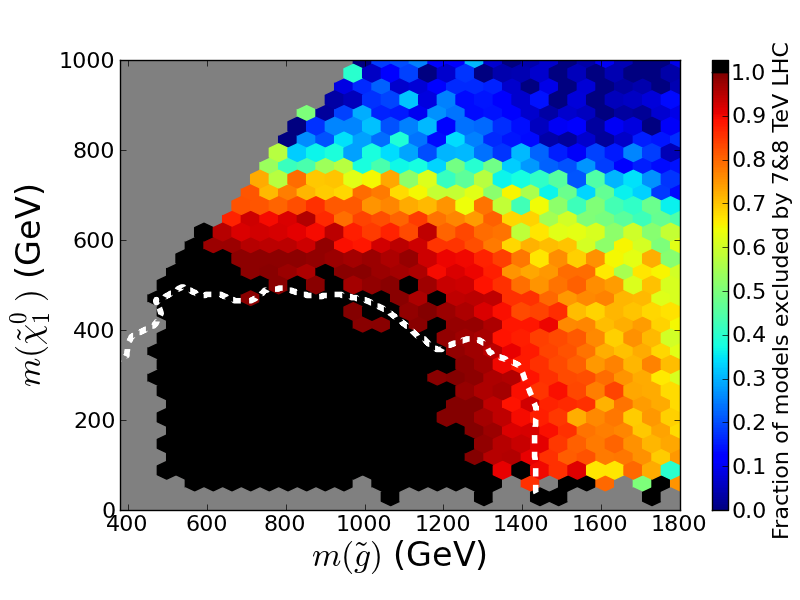}
\hspace{0.30cm}
\includegraphics[width=3.5in]{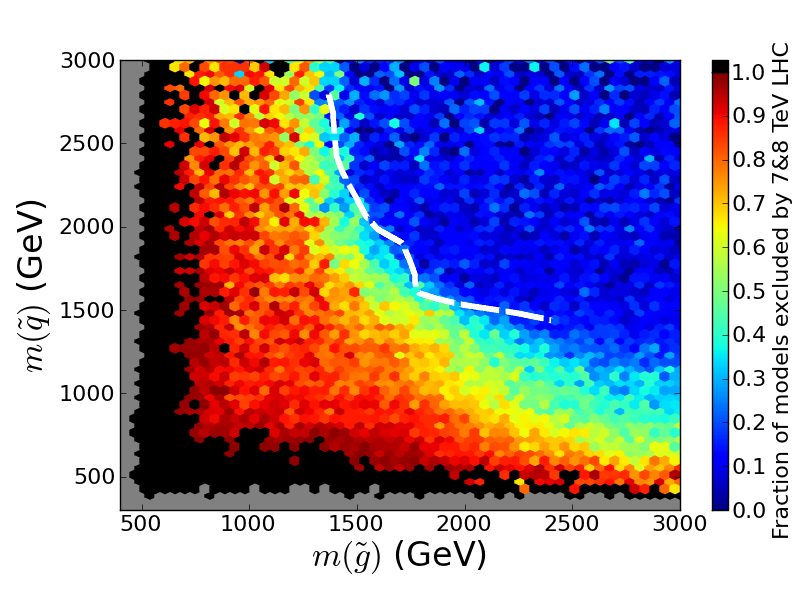}}
\vspace*{0.50cm}
\centerline{\includegraphics[width=3.5in]{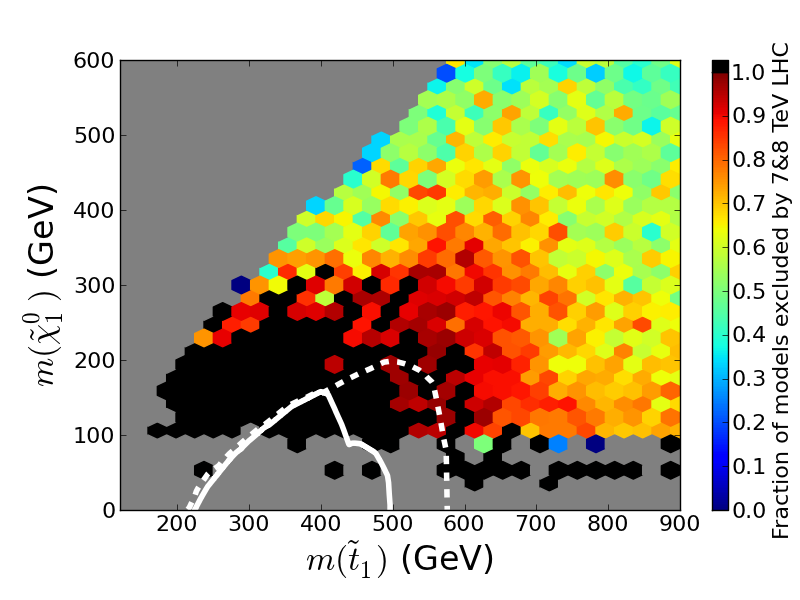}
\hspace{0.30cm}
\includegraphics[width=3.5in]{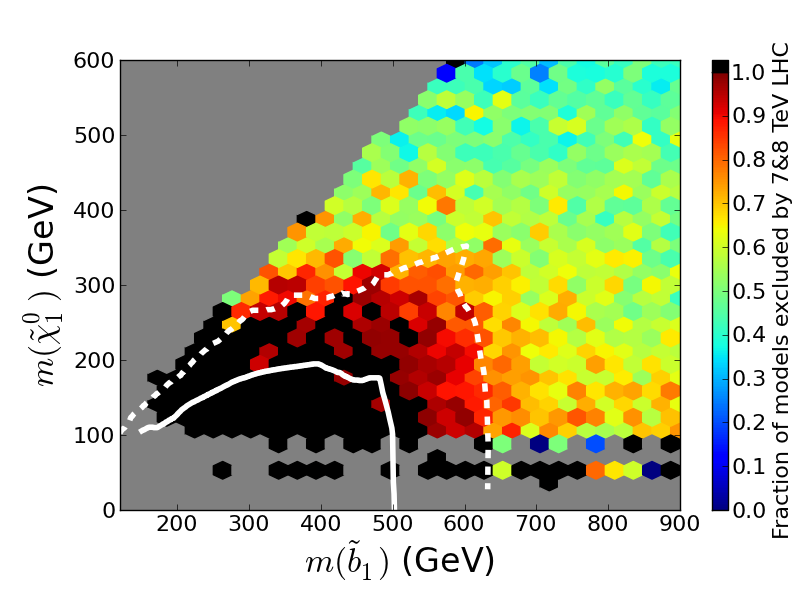}}
\vspace*{-0.10cm}
\caption{Projections of the pMSSM model exclusion efficiency from the 7 and 8 TeV LHC searches shown in the gluino-LSP mass plane (top left), 
the gluino-lightest squark mass plane (top right), the lightest stop-LSP mass plane (lower left) and the lightest sbottom-LSP mass plane (lower right). The 
solid and dashed lines represent the corresponding $95\%$ CL limit results at 7 and 8 TeV, respectively, obtained by ATLAS in the simplified model scenario as discussed in the text.}
\label{fig1}
\end{figure}

In addition to the 7 and 8 TeV LHC searches, future data taking and enhanced analyses at $\sim$ 14 TeV will greatly extend the expected coverage of the pMSSM parameter 
space. In~\cite{Cahill-Rowley:2013yla}, we considered the impact of one of the most powerful of these searches to be performed by ATLAS, the 
zero-lepton jets + MET final state at 14 TeV with both 300 fb$^{-1}$ and 3 ab$^{-1}$ of integrated luminosity~\cite {ATLAS-EP}}, following the procedure described above. Note that in extrapolating from 300 fb$^{-1}$ 
to 3 ab$^{-1}$, luminosity scaling has been employed to obtain the expected limit. As a result of limitations on CPU time, we generated 14 TeV results only for the $\sim 30.7$k 
neutralino LSP models that survived the 7 and 8 TeV LHC analyses and predict a Higgs mass of $126\pm 3$ GeV. We note that since the results of the 7 and 8 TeV analyses 
are essentially independent of the Higgs mass, it is quite likely that our results for this narrow Higgs mass range would in fact be applicable, to a very good 
approximation, to the entire neutralino LSP model set. We find that the 14 TeV jets+MET analysis with 300 (3000) fb$^{-1}$ of data is expected to exclude 90.7\% (97.1\%) of models which have the correct Higgs mass and survive the 7/8 TeV searches in Table~\ref{SearchList}.

To augment the 14 TeV jets + MET search, we have recently added a pair of signal regions in each of the zero- and one-lepton stop analyses as presented by ATLAS 
in~\cite{ATLAS-EP2}.  These analyses feature sliding missing energy and transverse mass cuts for optimal sensitivity to different stop masses, creating a large effective 
number of signal regions. We chose to examine the signal regions that were optimized for stop masses of 800 GeV and 1 TeV with 3 ab$^{-1}$ of integrated luminosity. We derived the $95\%$ $CL_s$ limits from the expected ATLAS background numbers and scaled these limits to estimate the sensitivity at 300 fb$^{-1}$ of integrated luminosity as well. 
Taken together, these signal regions are expected to exclude $\sim 17(44)\%$ of the neutralino models assuming an integrated luminosity of 0.3(3) ab$^{-1}$ at 14 TeV. 
Combining them with the zero-lepton, jets + MET search discussed above excludes 90.8\% (97.2\%) of the models surviving the 7/8 TeV searches in Table~\ref{SearchList}. This demonstrates that these additional signal regions are not expected to exclude very many of the models which are missed by the jets + MET search.

\begin{figure}[htbp]
\centerline{\includegraphics[width=3.5in]{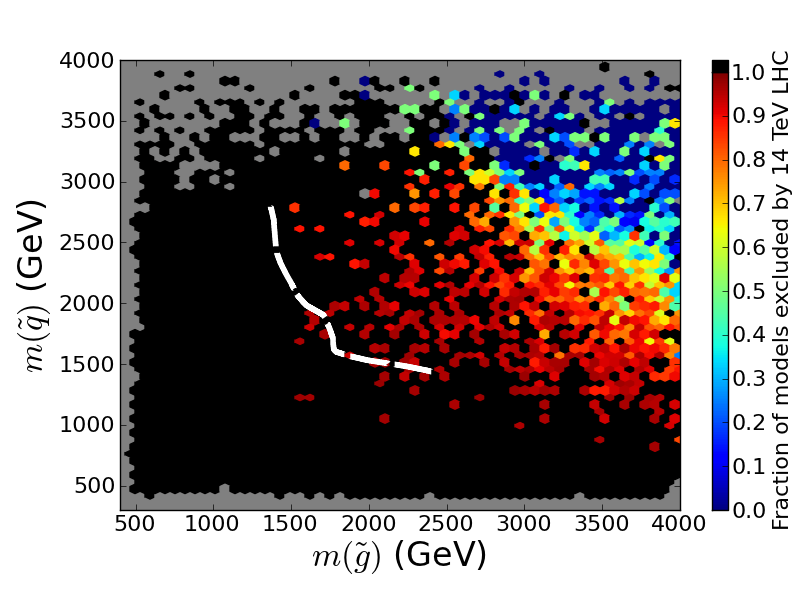}
\hspace{0.30cm}
\includegraphics[width=3.5in]{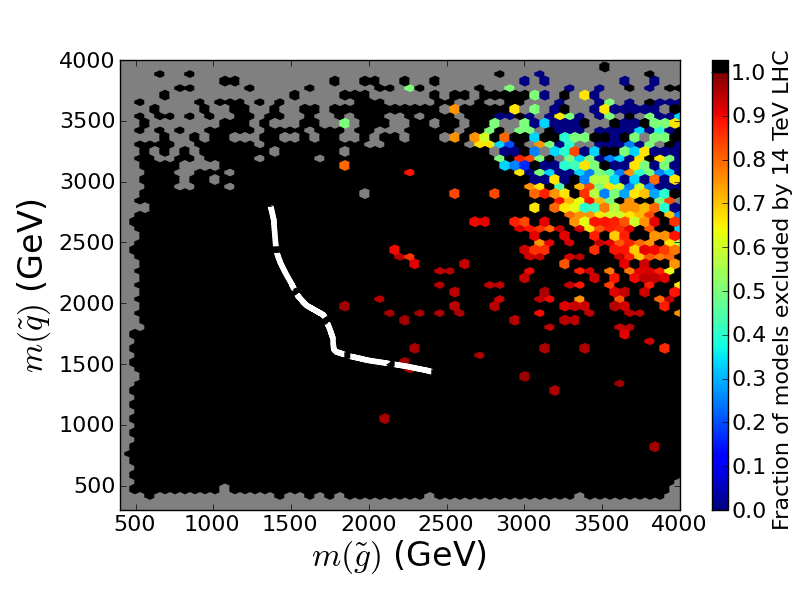}}
\vspace*{-0.10cm}
\caption{Fraction of models in the lightest squark-gluino mass plane which are expected to be excluded by combining our currently simulated 7/8 TeV searches with Jets+MET and stop searches at the 14 TeV LHC, assuming an integrated luminosity of 300 fb$^{-1}$ (left panel) and 3000 fb$^{-1}$ (right panel).}
\label{fig2}
\end{figure}

The exclusion reach of the combination of the ATLAS 14 TeV zero-lepton jets+MET search and the two stop searches is summarized in Fig.~\ref{fig2} for both 0.3 and 3 ab$^{-1}$ of 
integrated luminosity in the gluino-lightest squark mass plane. Here we see that for typical models, the 14 TeV LHC will be able to exclude squarks up to $\sim$ 1.6 TeV 
and gluinos up to $\sim$ 2.7 TeV. A few models are seen to survive with much lighter 
squark and/or gluino masses; in almost all cases these models survive by producing multiple high-$p_T$ leptons or b-jets from sbottoms (rather than stops) and therefore fall outside of the search regions. Adding additional searches with leptons in the signal region will undoubtedly exclude many of the surviving models with light colored sparticles. To fully understand the 
capabilities of the 14 TeV LHC will of course require a far more realistic study than is presently available since the LHC collaborations themselves are still unsure of 
how well their detectors will perform under the very high pileup conditions at 14 TeV.

\section{Direct Detection}

The direct detection of DM results from either the spin-independent (SI) or spin-dependent (SD) scattering of the LSP off of a target nucleus. While $Z(h)$ $t-$channel 
exchange only contributes to the SD(SI) process at tree level, $s-$channel squark exchange can contribute to both scattering processes. Clearly, as the bounds on the first 
and second generation squark masses from the LHC become stronger the importance of these squark exchange contributions will become sub dominant.{\footnote {It 
is important to note, however, that since the many different light squark masses can vary independently in the pMSSM, and since the LHC constraints on the $u_L,d_L$, 
$u_R$ and $d_R$ squarks are quite different, both SD and SI interactions may have substantial isospin-dependent contributions so that $\sigma_p$ and $\sigma_n$ can be 
significantly different.}} The $Z$-exchange graph is sensitive to the Higgsino content of the LSP, whereas the Higgs exchange graph probes the product of the LSP's gaugino 
and Higgsino content. Similarly, squark exchange is particularly sensitive to the LSP's wino and bino content. 

\begin{figure}[htbp]
\centerline{\includegraphics[width=4.0in]{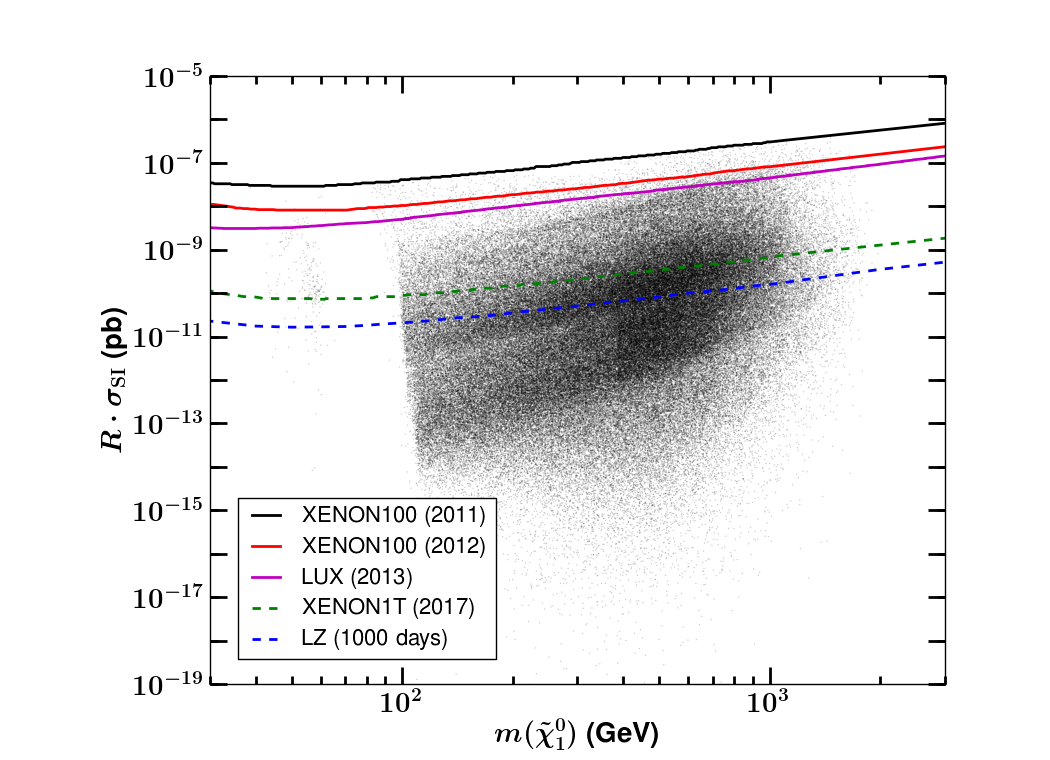}
\hspace{-0.50cm}
\includegraphics[width=4.0in]{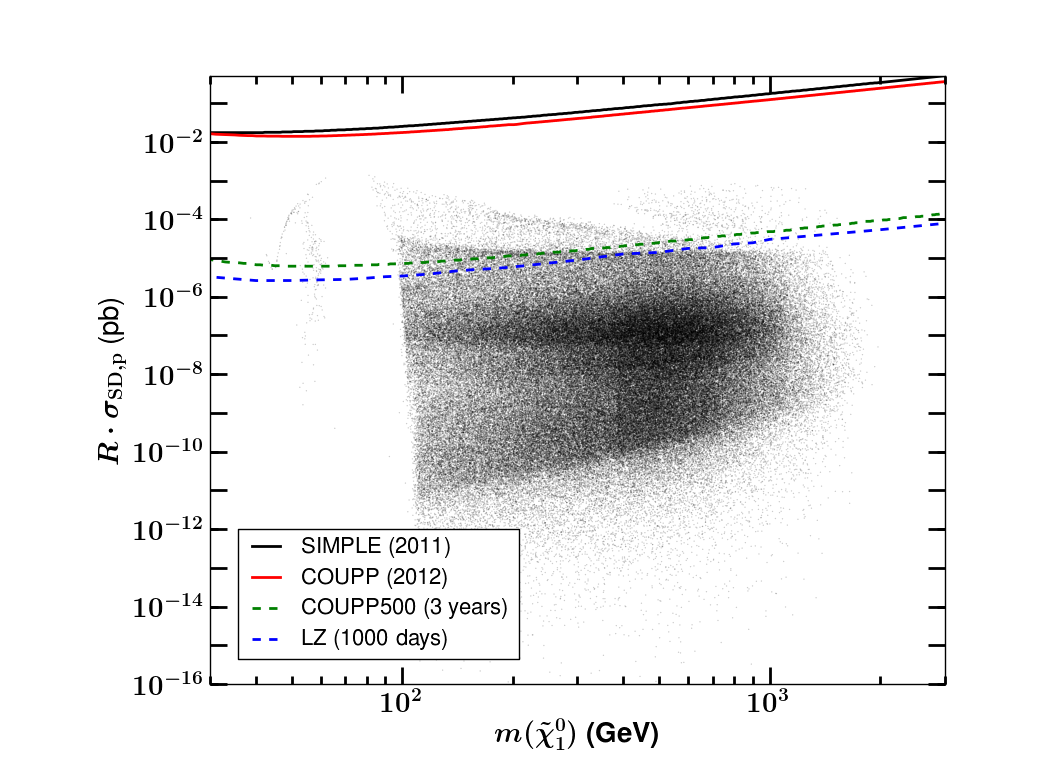}}
\vspace*{-0.10cm}
\caption{Scaled spin-independent (left) and spin-dependent (right) direct detection cross sections for our pMSSM model set in comparison to current 
and future experimental sensitivities. The scaling factor accounts for the possibility that the calculated thermal relic density of the LSP is below 
that measured by WMAP.}
\label{fig3}
\end{figure}

Figure~\ref{fig3} displays the predicted SI and SD cross sections for our pMSSM model set together with several present~\cite{Aprile:2011hi, Aprile:2012nq, Felizardo:2011uw, Behnke:2012ys} and anticipated 
future~\cite{Aprile:2012zx, COUPP500, LZ} experimental constraints. Note that the cross sections are appropriately scaled by the factor $R \equiv \Omega h^2/\Omega h^2|_{WMAP}$ to account 
for the fact that most of our pMSSM models lead to a thermal relic density somewhat below the WMAP value as discussed in the Introduction. 
There are two things to note in this Figure: ($i$) Future SI searches will cut rather deeply into the model set; a lack of signal at XENON1T(LZ) would 
exclude 23\%(39\%) of these models. {\it However}, this implies that more than half of our models are not accessible to SI experiments due to their 
rather small scaled cross sections, $R\sigma$. Models tend to produce these small values of $R\sigma$ due to both the suppression arising from 
their low thermal relic density as well as the tendency of the LSPs to be nearly pure weak eigenstates as discussed above. Note that if we only consider 
models which predict a relic density within 10\% of the critical density, the coverage improves significantly, with 60\% (81\%)) of models lying within the XENON1T (spin-independent LZ) expected search limit. ($ii$) SD searches are rather far from the pMSSM model predictions and we do not expect future SD reaches to have a significant impact on the 
parameter space: SD experiments such as COUPP500(LZ) will only be able to exclude $\sim 2(4)\%$ of the models in this set if no signal is observed.

\begin{figure}[htbp]
\centerline{\includegraphics[width=3.5in]{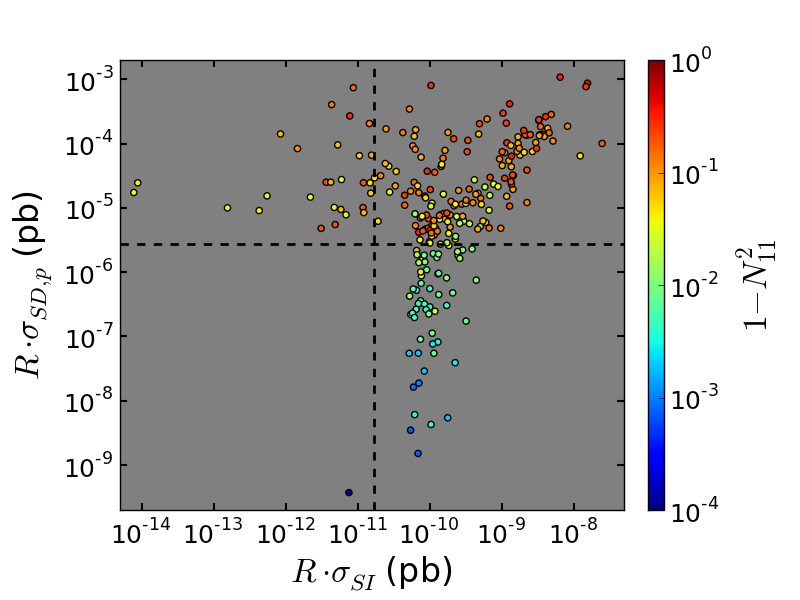}
\hspace{0.10cm}
\includegraphics[width=3.5in]{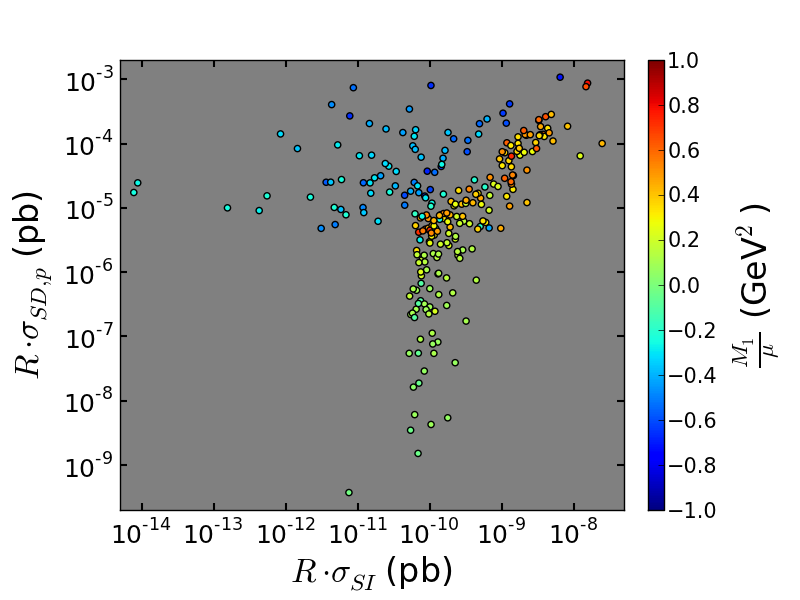}}
\vspace*{-0.10cm}
\caption{The correlation between spin-independent and proton spin-dependent scattering cross section for light LSPs. In the left panel, the points are colored according to their bino content. In the right panel, the color indicates the ratio $M_1/\mu$, showing the effect of the relative sign between $M_1$ and $\mu$ on the cancellation of Higgs exchange contributions to spin-independent direct detection.}
\label{fig:sisd}
\end{figure}

Direct detection experiments can apply strong constraints on specific LSP compositions or annihilation mechanisms. An interesting example is the case of an LSP with a mass significantly below the LEP limit on chared sparticles. This LSP is required to be mostly bino (otherwise it would be accompanined by an excluded chargino), and is generally prevented from coannihilating by the LEP limit on charged sfermions (although in some cases sneutrino co-annihilation may be possible). The dominant annihilation mode is therefore through s-channel Z or Higgs bosons. The left panel of Figure~\ref{fig:sisd} shows these light LSPs in our model set ($m_{LSP} < 80$ GeV) in the scaled SI vs scaled SD cross section plane; we find that while many of the models have a SI or SD cross-section beyond the reach of current or future experiments, only one model (where the LSP co-annihilates with a 100 GeV stau) is expected to remain undetected by the combination of future SI and SD searches at LZ. Naively, one would not expect such complementarity between the SI and SD scattering experiments for light bino LSPs with small Higgsino components, since both cross sections should both be determined by the Higgsino content of the LSP. However, we see from the left panel of the figure (where the points are color-coded according to their bino/Higgsino content with $N_{11}^2$ denoting the bino fraction which approaches unity for a 100\% bino eigenstate) that this expectation is realized for the SD cross section but not for the SI cross section. In particular, very small SI cross-section values are obtained for large Higgsino content. The suppression of the SI cross section despite large Higgsino content results from cancellation between heavy and light Higgs exchange diagrams; this can be seen in the right panel of the figure, which colors the points according to the value of the ratio $M_1/\mu$. The sign of this quantity determines the relative sign of the mixing matrix elements; when it is negative, a relative sign appears between the light Higgs coupling and the heavy neutral Higgs couplings, allowing for cancellation between the light and heavy Higgs exchange contributions. Although the heavy Higgs bosons in our model set are frequently very massive, the cancellation can still occur since the dominant contribution to scattering comes from Higgs exchange with a strange quark, which is $\tan{\beta}$ enhanced for heavy Higgs bosons in the decoupling limit. Additional cancellations with squark diagrams can lead to further suppression of the SI cross section. Interestingly, a negative value of the ratio $M_1/\mu$ also tends to result in a smaller coupling to the light Higgs, with the result that a larger Higgsino fraction (and therefore a larger SD cross-section) is required in order to provide a sufficiently large annihilation cross-section. 

While direct detection experiments have significant power to cover much of the pMSSM, clearly such experiments will need to be supplemented if we want to 
discover or exclude the full range of neutralino LSPs in the pMSSM model space. In the next two sections, we describe constraints on the pMSSM from indirect detection and neutrino telescope experiments.

\section{Indirect Detection: \Fermi ~LAT and CTA}

Indirect detection plays a critical role in searches for DM and, in
the case of null results, can lead to very strong constraints on the
pMSSM parameter space. As will be seen below, both \Fermi ~and CTA can
contribute in different regions of the pMSSM parameter space in the
future.  CTA, in particular, will be seen to be extremely powerful in
the search for heavy LSPs which are mostly Higgsino- or wino-like and
that predict thermal relic densities within an order of magnitude of
the WMAP/Planck value. \Fermi, ~on the other hand, will be seen to be
mostly sensitive to well-tempered neutralinos that are relatively
light.

The most promising DM targets for both \Fermi ~LAT and CTA are those
with both a high DM density and low astrophysical gamma-ray
foregrounds.  These criteria have motivated a number of Galactic and
extragalactic targets including the Galactic Center (GC), dwarf
spheroidal galaxies (dSphs), and galaxy clusters. The expected
gamma-ray signal for DM annihilations is proportional to the integral
of the square of the DM density along the line of sight to the source
($J$).  The determination of $J$ is most reliable for DM-dominated
objects such as dSphs and galaxy clusters in which the DM distribution
can be robustly measured.
In the Milky Way halo the uncertainty on the DM distribution rapidly
increases as one approaches the inner galaxy where baryons dominate
the gravitational potential.  Kinematic data at large scales constrain
the density of DM at the solar radius to 0.2--0.4~GeV~cm$^{-3}$
\cite{Catena:2009mf, Bovy:2012tw}.  

Under well-motivated models for the DM distribution in the Galactic
halo, the GC is expected to be the most intense DM source in the sky.
CDM simulations predict that the Galactic DM halo should have a
density profile with an inner cusp, $\rho \propto r^{-\gamma}$ with
$\gamma \simeq 1.0$.  For an extrapolation of the Galactic DM density
profile with $\rho(r) = \rho_{\odot}(r/r_\odot)^{-1}$, the expected DM
signal from the GC region is approximately two orders of magnitude
greater than dSphs or galaxy clusters.  However the GC is also the
region of the sky with the highest density of gamma-ray sources and
brightest diffuse gamma-ray emission produced from the interaction of
cosmic rays with the interstellar medium.  These foregrounds
significantly limit the sensitivity of the \Fermi-LAT in the inner
galaxy and complicate the interpretation of any observed signals.  

For our study of \Fermi-LAT we focus on the sensitivity to the DM
signal from dSphs which are predominantly at high galactic latitudes
where the astrophysical foregrounds are much weaker.  Above 50~GeV the
diffuse emission from the Galaxy is much less intense relative to
other backgrounds than in the \Fermi-LAT sensitivity band.  The inner
galaxy is thus the preferred target for CTA under the assumption that
the Galactic DM halo possesses an inner cusp with $\gamma \simeq 1.0$.
When considering signals from models with a relic density below the
WMAP value, we rescale the annihilation cross section by $R^{2}$ to
account for the reduced number density of DM particles and consequent
reduction in the $J$ factor.  Implicit in this rescaling is that the
LSPs in these models constitute only one component of DM.

\subsection{\Fermi ~LAT}

Here we follow the procedure developed in Ackermann et al.~(henceforth A11) \cite{Ackermann:2011wa} and expanded upon in Cotta et al.~\cite{Cotta:2011pm} and Ackermann et al.~\cite{Ackermann:2014yva} to constrain the annihilation cross section, \sigv, for each pMSSM model using \Fermi-LAT observations of ten dwarf spheroidal galaxies (dSphs).  
Our two-year $\gamma$-ray event sample is identical to that described in A11, accepting photons in the energy range from $200\,\MeV < E < 100\,\GeV$ within 10\degree of each dSph. 
In accord with A11, we use the LAT ScienceTools\footnote{\url{http://fermi.gsfc.nasa.gov/ssc/data/analysis/software}} version v9r20p0 and the P6\_V3\_DIFFUSE instrument response functions.\footnote{\url{http://fermi.gsfc.nasa.gov/ssc/data/analysis/scitools/overview.html}} 
J-factors and associated statistical uncertainties for the dSphs are taken from Table 1 of A11, where they were calculated using line-of-sight stellar velocities and the Jeans equation \cite{Ackermann:2011wa}. 
Similar to Cotta et al., we use DarkSUSY 5.0.5 \cite{Gondolo:2004sc} to model the $\gamma$-ray spectrum from the annihilation of each pMSSM LSP. 
DarkSUSY calculates the total $\gamma$-ray yield from annihilation, as well as the rates into each of 27 final state channels. 

We create bin-by-bin likelihood functions from the Pass 6 data surrounding each of the 10 dwarf spheroidal galaxies following the procedure of Ackermann et al.~\cite{Ackermann:2014yva}.
We calculate a joint likelihood to constrain the annihilation cross section of each pMSSM model given the LAT observations coincident with the ten dSphs.
Following A11, we incorporated statistical uncertainties in the J-factors of the dSphs as nuisance parameters in our likelihood formulation~(see Equation 1 and the associated discussion in A11). 
This likelihood formulation includes both the flux normalizations of background $\gamma$-ray sources (diffuse and point-like) and the associated dSph J-factors and statistical uncertainties. 
No significant $\gamma$-ray signal is detected from any of the dSphs when analyzed individually or jointly for any of the pMSSM models.
Utilizing the publicly available bin-by-bin likelihood functions derived from the analysis of Pass 7 rather than a re-analysis of the Pass 6 data would not qualitatively alter these results.

For each of the pMSSM models, we calculate the maximum annihilation cross section, \sigvm, consistent with the null detection in the LAT data.
We incorporate nuisance parameters to obtain a 95\% one-sided confidence interval on the value of \sigv using the profile likelihood method~\cite{Rolke:2004mj}. 
This one-sided 95\% confidence limit on \sigv serves as our value of \sigvm, which is compared to the true value of the annihilation cross section for each pMSSM model. 
We define the ``boost'' necessary to constrain a model as the ratio $\sigvm/\sigv$.

While the LAT data do not presently constrain any of the pMSSM models, it is useful to estimate the improvements expected over a 10 year mission lifetime. 
In the low-energy, background dominated regime, the LAT point source sensitivity increases as roughly the square-root of the integration time.
However, in the high-energy, limited background regime (where many pMSSM models contribute), the LAT sensitivity increases more linearly with integration time. 
Thus, 10 years of data could provide a factor of $\sqrt{5}$ to 5 increase in sensitivity. Additionally, optical surveys such as Pan-STARRS~\cite{Kaiser:2002zz}, the Dark Energy Survey~\cite{Abbott:2005bi}, and LSST~\cite{Ivezic:2008fe} could provide a factor of 3 increase in the number of Milky Way dSphs corresponding to an increased constraining power of $\sqrt{3}$ to 3~\cite{Tollerud:2008ze}.  
Ongoing improvements in LAT event reconstruction, a better understanding of background contamination, and an increased energy range are all expected to provide additional increases in the LAT sensitivity. 
Thus, we find it plausible that the LAT constraints could improve by a factor of 10 compared to current constraints. 

\begin{figure}[htbp]
\centerline{\includegraphics[width=3.5in]{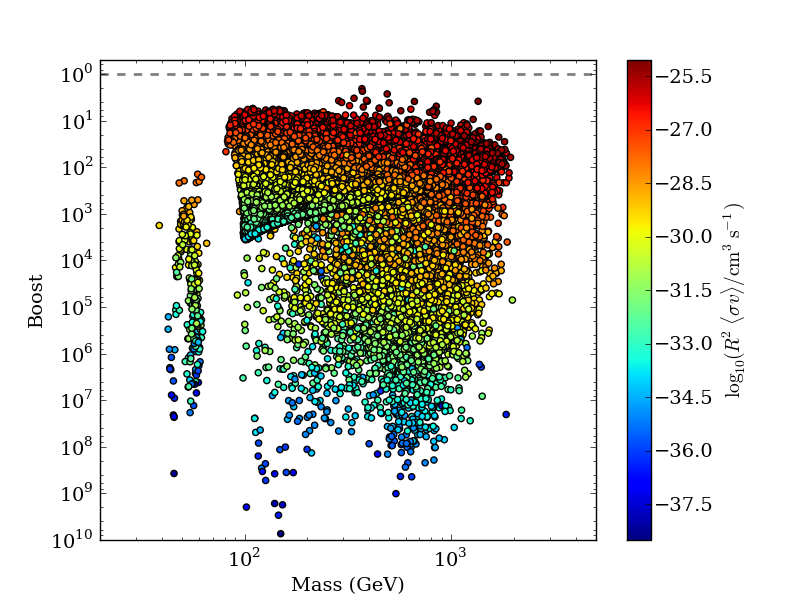}
\hspace{-0.50cm}
\includegraphics[width=3.5in]{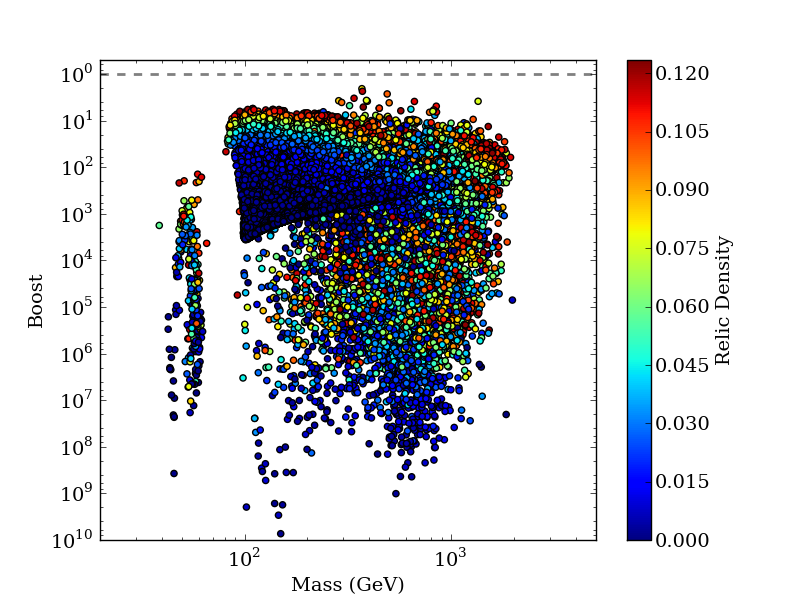}}
\vspace*{-0.10cm}
\caption{(Left) \Fermi-LAT boost factor vs. LSP mass for the pMSSM model set. The total cross section scaled by the DM
    fraction, $R^{2}$\sigv, for each model is plotted on the 
color scale. (Right) Here the corresponding relic density for each model is plotted on the color scale.}
\label{fig4}
\end{figure}

In Figure~\ref{fig4} we display the boost required to constrain the various pMSSM models at $95\%$ CL based on the \Fermi-LAT dwarf analysis employing only the 
first 2 years of data color-coded by either the annihilation cross section or the LSP thermal relic density. 
Here we see that the LAT analysis does not currently constrain any of our pMSSM models. 
However, as discussed above with more dwarfs and longer integration times we would expect an $\sim 10$-fold improvement in the sensitivity and thus all models with boost factors less than 10 would become accessible. 
We will assume this $\sim 10$-fold improvement in sensitivity for the analysis that follows.

\subsection{CTA}

The Cherenkov Telescope Array (CTA) \cite{Consortium:2010bc} is a
future ground-based gamma-ray observatory that will have sensitivity
over the energy range from a few tens of GeV to a few hundreds of TeV.
To achieve the best sensitivity over this wide energy range CTA will
include three telescope types: Large Size Telescope (LST, 23 m
diameter), Medium Size Telescope (MST, 10-12 m) and Small Size
Telescope (SST, 4-6 m).  Over this energy range the point-source
sensitivity of CTA will be at least one order of magnitude better than
current generation imaging atmospheric Cherenkov telescopes such as
H.E.S.S., MAGIC, and VERITAS.  CTA will also have an angular
resolution at least 2--3 times better than current ground-based
instruments, improving with energy from 0.1$^\circ$ at 100 GeV to
better than 0.03$^\circ$ at energies above 1 TeV.

The optimal DM search region for CTA will be limited by the CTA FoV of
$\sim$8$^\circ$ to the area within $2^\circ$ to $3^\circ$ of the GC.
The DM signal on these angular scales predominantly probes the
DM distribution in the inner galaxy ($R_{GC} <1$~kpc).  We model the
Galactic DM distribution with an NFW profile with a scale radius of 20
kpc normalized to 0.4~GeV~cm$^{-3}$ at the solar radius.  This model
is consistent with all current observational constraints on the
Galactic DM halo and represents a conservative expectation for the
inner DM profile in the absence of baryonic effects.  

Because the annihilation signal is proportional to the square of the
DM density, the projected limits for CTA depend strongly on the
assumptions that are made on the shape and normalization of the
Galactic DM halo profile.  The projected limits presented here could
change by as much as a factor of 10 given these uncertainties.  The
analysis strategy adopted for this study also relies on the existence
of a cusp in the MW DM density profile which would produce a
measurable gradient within the FoV of CTA.  MW density profiles with a
central core would require a different analysis strategy than the one
presented here and would likely result in a reduced sensitivity to a
DM signal in the GC.

The prospects for CTA to detect DM and test other exotic physics has
been studied in detail by \cite{Doro:2012xx} using models for the CTA
response functions from \cite{Bernlohr:2012we}.  These models were
derived from detailed Monte Carlo simulations generated for a variety
of possible array configurations with 18--37 MSTs and different
combinations of SSTs and LSTs.  The baseline design for CTA is a
balanced array with 18--25 MSTs, 3--4 LSTs, and 50--70 SSTs that
maximizes the performance over the whole CTA energy range.

For this study we model the performance of CTA using simulations of an
array with 61 MSTs distributed on a regular grid with 120~m spacing
\cite{Jogler:2012}.  For a gamma-ray source with the spectral
properties of DM, the sensitivity of this array is similar to the
expected sensitivity of the baseline CTA design with a US extension of
$\sim$24 MSTs.
The array used for this study has a gamma-ray angular resolution that
can be parameterized as a function of energy as $\theta \simeq
0.07^\circ(E/100$~GeV)$^{-0.5}$ and a total gamma-ray effective area
above 100 GeV of $\sim$10$^{6}$ m$^{2}$.  We define the GC signal
region as an annulus centered on the GC that extends from
0.3$^\circ$--1.0$^\circ$ and calculate the sensitivity of CTA for an
integrated exposure of 500 hours that is uniform over the whole
region.  An energy-dependent model for the background in the signal
region is taken from a simulation of residual hadronic contamination.
The uncertainty in the background model is calculated for a control
region with no signal contamination and a solid angle equal to five
times the signal region (14.3~deg$^{2}$).

We estimate the sensitivity of CTA using a binned likelihood analysis
with two model components: an isotropic background that models the
distribution of residual cosmic rays and a template for the DM
annihilation signal.  The likelihood of the signal and background
components is evaluated from the distribution of events in a
two-dimensional map binned in energy and angular offset from the GC.
For each model, the maximum cross section consistent with a null
detection at the 95\% C.L. ($\sigvm$) is calculated from the ratio
between likelihoods evaluated with and without the DM component.
Following the same procedure as the \Fermi-LAT analysis, we compute
the model boost factor as the ratio of the model cross section with
$\sigvm$.

\begin{figure}
  
  \centering
  \includegraphics[width=0.49\textwidth]{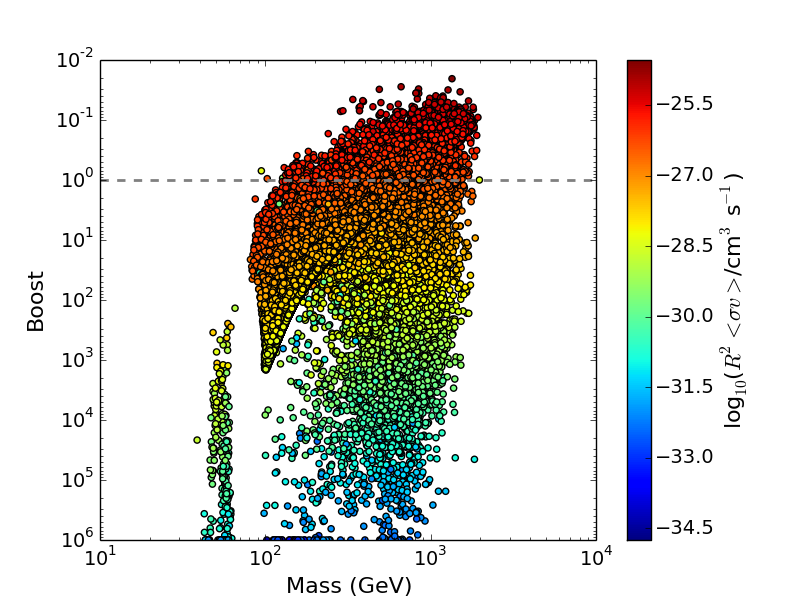}
  \includegraphics[width=0.49\textwidth]{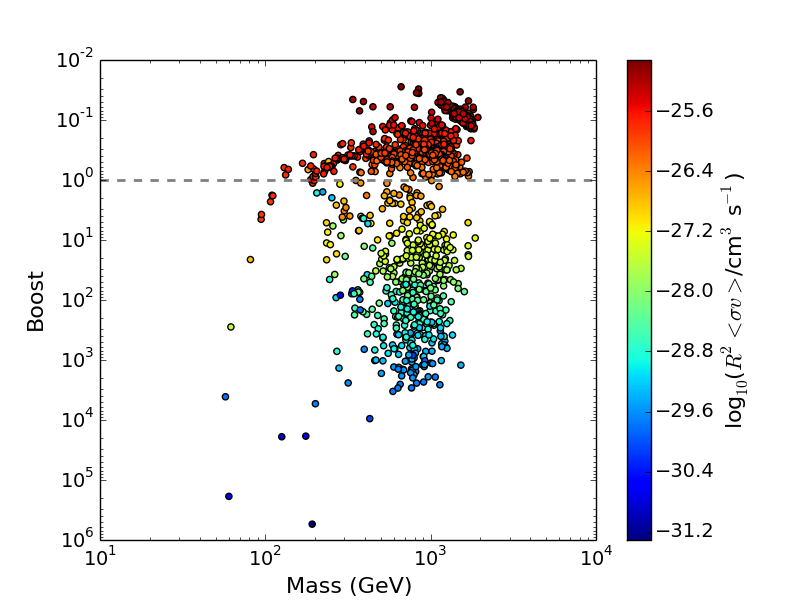}
  \caption{\label{fig:cta_pmssm_sigmav}CTA boost factor vs. LSP mass
    for the full pMSSM model set (left) and pMSSM models that survive
    7+8 TeV LHC searches and have a LSP relic density consistent with
    the WMAP/Planck value. The total cross section scaled by the DM
    fraction, $R^{2}$\sigv, is plotted on the color scale.  Models
    above the dashed grey line would be excluded by CTA at the 95\%
    C.L. given an exposure of 500 hours on the GC.}
\end{figure}

Figure \ref{fig:cta_pmssm_sigmav} shows the distribution of the CTA
boost factor versus LSP mass for all pMSSM models and the subset of
models that have an LSP relic density consistent with 100\% of the DM
relic density.  For models with LSP masses above 100--200~GeV, the
sensitivity of CTA is observed to be well correlated with the total
annihilation cross section.  At lower LSP masses, the boost factor
distribution begins to shift to higher values as the peak of the
gamma-ray annihilation spectrum moves below the energy threshold of
CTA ($\sim30$~GeV).  CTA can exclude $\sim$20\% of the total model set
and $>$50\% of the models in the subset of models with an LSP relic
density that saturates the WMAP/Planck value.  In both scenarios the
majority of models excluded by CTA are those which have a pure wino or
Higgsino LSP with a mass near 1~TeV.

\section{IceCube}

Neutralino dark matter can be captured and accumulated in the sun. Neutralinos in this relatively over dense population would then sink to the 
solar core and annihilate. If the product of capture and annihilation cross sections is large enough this process leads to an equilibrium
 population of captured neutralinos whose annihilations are proportional to their \emph{elastic scattering cross-sections} 
\cite{Press:1985ug,Gould:1987ir} and that may be 
detectable by observing an excess of high-energy ($\geq \rm{GeV}$) solar neutrinos in km$^3$-scale neutrino telescope experiments~\cite{Silverwood:2012tp}.

Here we present predictions for an IceCube/DeepCore (IC/DC) search for neutralino DM in our pMSSM model set. Our analysis closely 
follows that presented in \cite{Cotta:2011ht}. In
the results presented here we assume that each neutralino's relic density is given by the usual thermal calculation. We use 
DarkSUSY 5.0.6 \cite{Gondolo:2004sc} to simulate the 
(yearly-average) signal $\nu_{\mu}$/$\bar{\nu}_{\mu}$ neutrino flux spectra incident at the detector's position and convolve with 
preliminary
 $\nu_{\mu}$/$\bar{\nu}_{\mu}$ effective areas for muon events contained in DeepCore\footnote{These are the same effective areas that 
were used in \cite{Cotta:2011ht},
 referred to there as ``SMT8/SMT4."}. We consider a data set that includes $\sim 5$yr of data that is taken during austral winters 
(the part of the year for
 which the sun is in the northern hemisphere) over a total period of $\sim 10$yrs\;\footnote{In practice, the IC/DC treatment of data 
is more sophisticated, classifying
events as through-going, contained and strongly-contained, and allowing for some contribution from data taken in the austral summer. 
We expect that inclusion of this data
 would affect our results at a quantitative, but not qualitative, level.}. An irreducible background rate of $\sim 10$ events/yr is 
expected from
cosmic ray interactions with nuclei in the sun. Here we will take (as discussed at greater length in \cite{Cotta:2011ht}) a detected 
flux of $\Phi=40$ events/yr as a conservative
 criterion for exclusion. 

\begin{figure}[hbtp]
    \centering
    \includegraphics[width=1.0\textwidth]{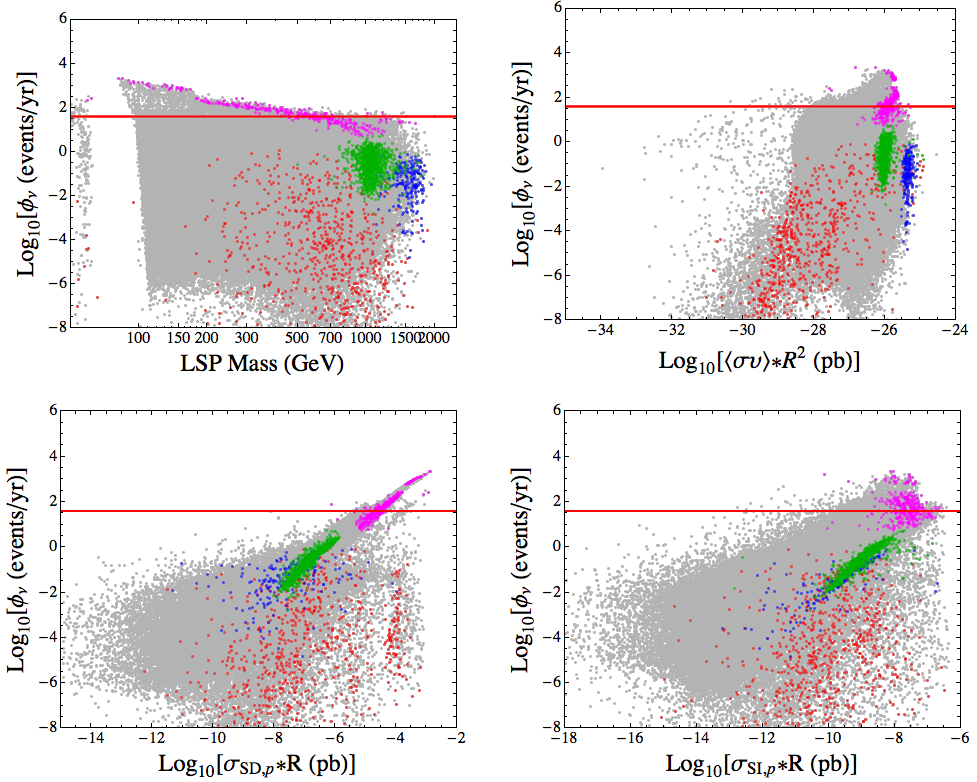}
    \caption{IC/DC signal event rates as a function of LSP mass (upper-left), scaled thermal annihilation cross-section 
      $\langle\sigma\upsilon\rangle R^2$ (upper-right) and scaled thermal elastic scattering cross-sections 
      $R \times \sigma_{SD,p}$ and $R \times \sigma_{SI,p}$ (lower panels). In all panels the gray points represent the
      models in our full pMSSM sample, while WMAP-saturating models with mostly bino, wino, Higgsino 
      or mixed ($\leq$80\% of each) LSPs in are highlighted in red, blue, green and magenta, respectively.
      The red line denotes a detected flux of $40$ events/yr, our conservative estimate for exclusion.}
    \label{icdc}
  \end{figure}

The basic results of this analysis are presented in Figure \ref{icdc}. In this figure, the full pMSSM model set is depicted by the gray points, and the WMAP-saturating models with mostly bino, wino, Higgsino or mixed ($\leq$80\% of each) LSPs are highlighted in red, blue, green and magenta,
 respectively. Detectability is tightly correlated with the elastic scattering cross-sections ($\sigma_{SI}$ and $\sigma_{SD}$) while having
 little correlation with the annihilation cross-section $\langle\sigma\upsilon\rangle$, as expected. 

The biggest difference between these results and those of the previous analysis \cite{Cotta:2011ht}, which used an older set of pMSSM models that
 were chosen to have relatively light ($\leq 1\tev$) sparticles, is that a much smaller percentage of the current pMSSM models are able to reach
 capture/annihilation equilibrium in the sun. This is due to the fact that so many of these models are nearly pure wino or Higgsino gauge 
eigenstates (which have both low
relic density and small capture cross-sections) and that the LSPs in this model set tend to be much heavier than those in the previous set. 
If one defines
out-of-equilibrium models as those with solar annihilation rates less than 90\% of their capture rates, we find that no such models can be 
excluded by IC/DC. In contrast,
 relatively light LSPs composed of a mixture of gaugino and Higgsino eigenstates have large scattering and annihilation cross sections and 
are highly detectable by IC/DC.
 We observe that all such WMAP-saturating well-tempered neutralinos with masses $m_{LSP}\leq 500\gev$ should be excluded by the 
IC/DC search (\emph{c.f.}, the magenta
 points in Fig. \ref{icdc}).

\section{Complementarity: Putting It All Together}

Now that we have provided an overview of the various dark matter searches that form our analysis, we can combine them to see what they (will) reveal about the 
nature of the neutralino LSP as DM~\cite{Boehm:2013qva} and, more generally, the pMSSM itself. Since we only have 14 TeV results for the $\sim$ 30.7k neutralino 
LSP models that survive the 7 + 8 TeV searches and have $m_h=126\pm 3$ GeV (because of CPU limitations as described above), the main results presented below will 
only make use of the 7 + 8 TeV LHC searches listed in Table~\ref{SearchList}. We will also present some indicative results showing the sensitivity of the combined 7, 8, and 14 
TeV LHC analyses for the subset of neutralino LSP models with $m_h=126\pm 3$ GeV.

\begin{figure}[htbp]
\centerline{\includegraphics[width=3.5in]{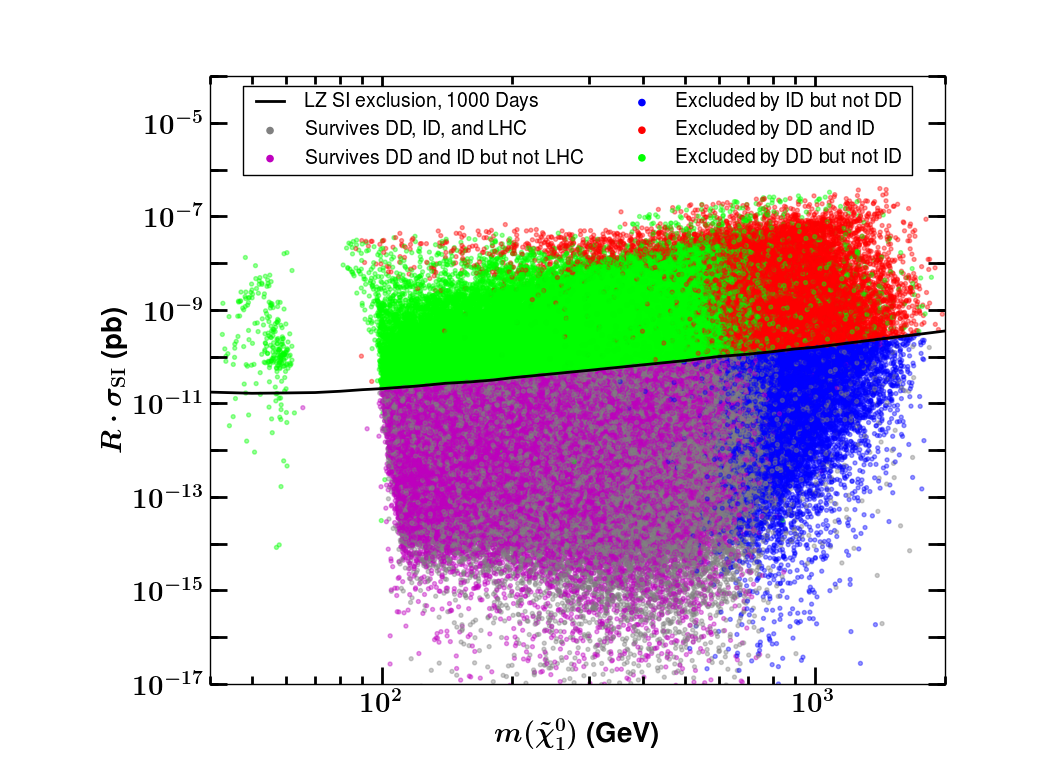}
\hspace{-0.50cm}
\includegraphics[width=3.5in]{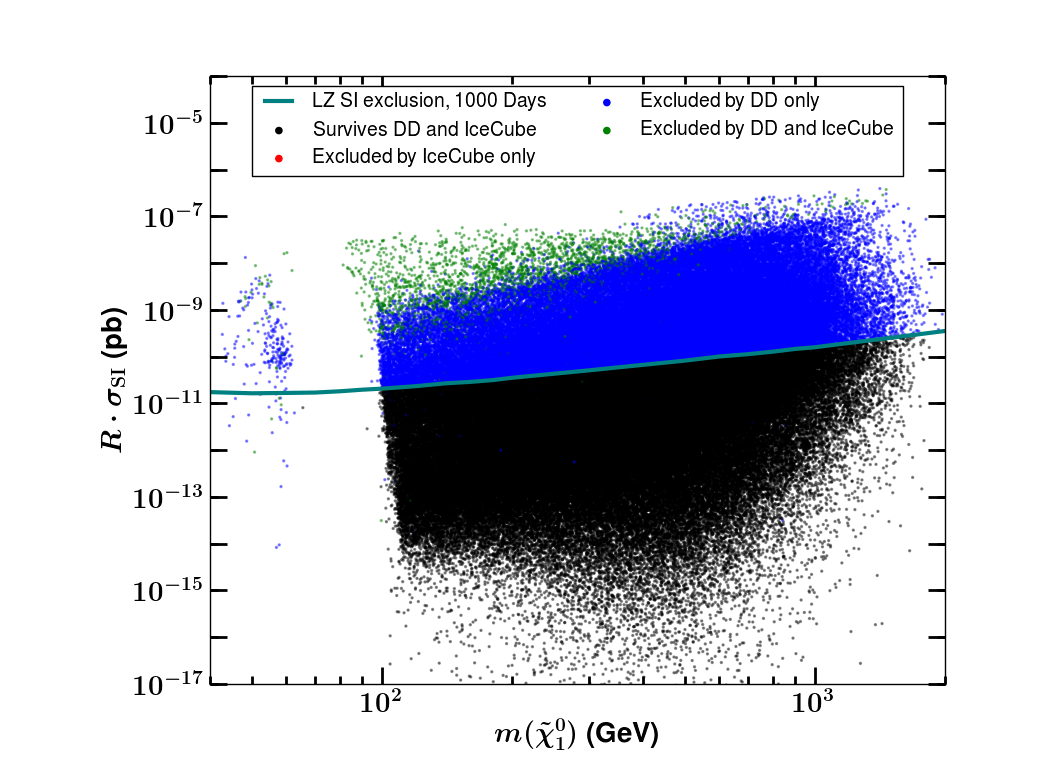}}
\vspace*{0.50cm}

\centerline{\hspace{0.3 cm} \includegraphics[width=3.5in]{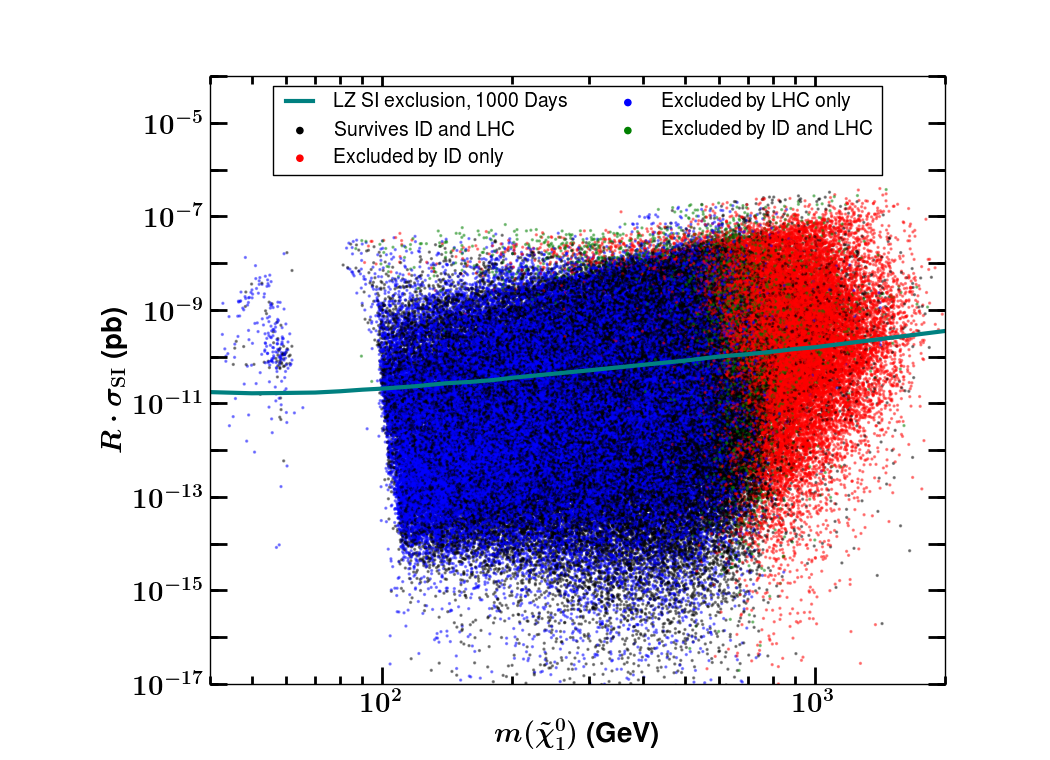}
\hspace{-0.05 cm}
\raisebox{-0.3 cm}{\includegraphics[width=3.5in]{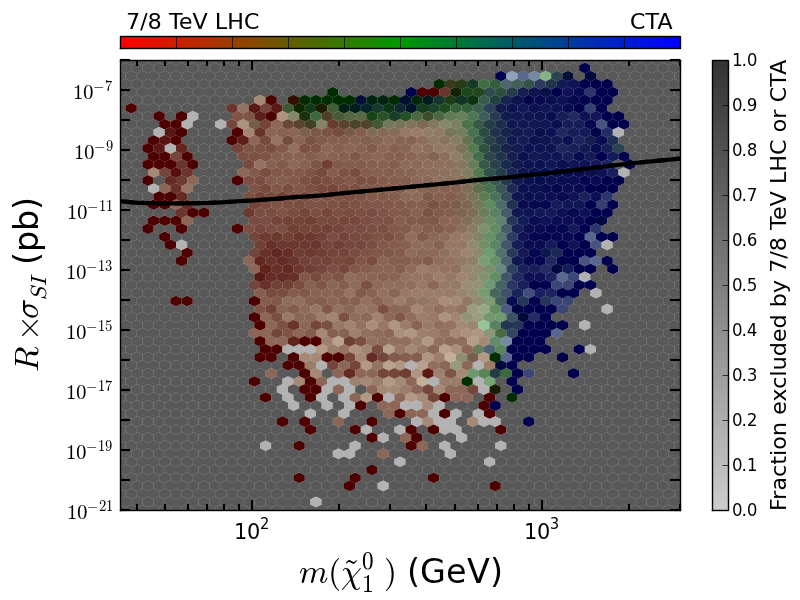}}}
\vspace*{-0.10cm}
\caption{Comparisons of the sensitivity of the various searches, color-coded as indicated, in the LSP mass-scaled SI cross section plane for the pMSSM model sample as discussed 
in the text. The anticipated SI limit from LZ is shown as a guide to the eye.}
\label{figxx}
\end{figure}

Figure~\ref{figxx} shows the survival and exclusion rates resulting from the various searches and their combinations in the LSP mass-scaled SI cross section plane. In 
the upper left panel we compare the combined direct detection (DD = LZ, SI + SD) and indirect detection (ID = \Fermi ~+ CTA) DM searches. Here we see that
7.2\% (27.5\%) of the models can be excluded by ID but not DD (excluded by DD but not ID) while 11.4\% are excluded by both types of searches. On the other hand, we also see that 
53.9\% of the models survive both sets of DM searches; 45.5\% of this subset of models, in turn, are presently excluded by the LHC. Note that the DD and ID searchable regions are relatively well separated in terms of mass and cross section although there is some overlap between the sets of models covered by the different 
experiments. In particular we see that the ID searches (here almost entirely CTA) are covering the heavy LSP region even in cases where the SI cross section is very 
low and likely beyond the reach of any potential DD experiment. Combining all of the searches only 24.7\% of the model set would remain undetected. 
Similarly, the upper right panel compares the reach of IceCube with DD and we see that 37.7\% (0\%) of the models are covered uniquely by DD (IceCube) only while 1.2\% can be simultaneously excluded by both 
sets of searches and 61.1\% would be missed by either search. 
In the lower left panel, ID and LHC searches are compared and we see that 15.7\% (42.4\%) of the models would be excluded only by the ID (LHC) searches. However, 
3.0\% (38.9\%) of the models are seen to be covered by (would be missed by) both search techniques. The strong complementarity between the LHC, CTA and LZ experiments 
is evident here as CTA probes the high LSP mass region very well where winos and Higgsinos dominate, LZ chops off the top of the distribution where the well-tempered 
neutralino LSP states dominate, and the LHC covers the relatively light LSP region (fairly independent of LSP type) rather well. Of course the strength of the LHC 
coverage will significantly improve when the 14 TeV analyses are included, as we will see below. In the lower right panel, the relative contributions arising from the LHC 
and CTA searches to the model coverage are shown. Here the color intensity of a given bin indicates the fraction of models in that bin excluded by the combination of both 
CTA and the LHC, while the hue indicates whether the excluded models are seen mostly by CTA (blue) or by the LHC (red). It is again quite clear that CTA completely dominates for large LSP masses and also competes with the LHC throughout the band along the top of the distribution, which mostly contains models with thermal relic densities approximately saturating the WMAP/Planck limit. The LHC is seen to exclude a significant fraction of models with LSP masses below $\sim$ 700 GeV, although there is no region in which the LHC excludes as large a fraction of models as CTA for the high LSP masses.

\begin{figure}[htbp]
\centerline{\includegraphics[width=3.5in]{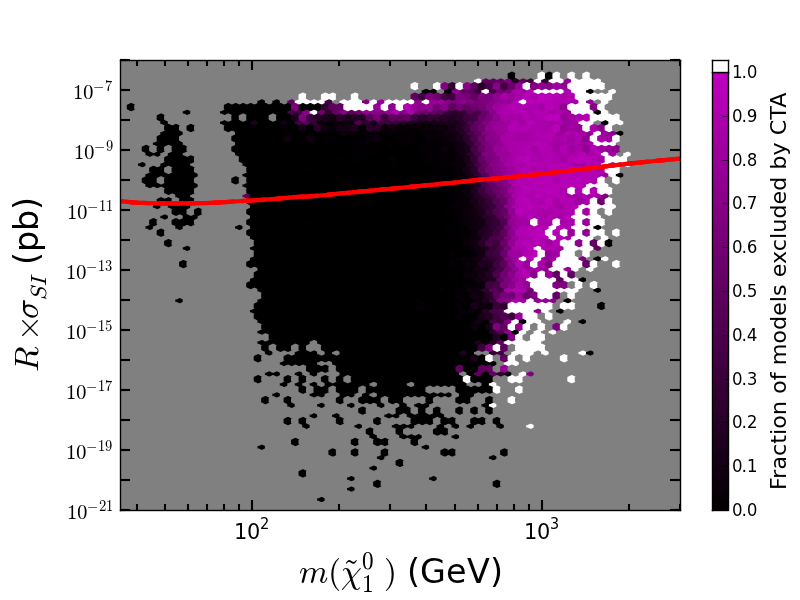}
\hspace{0.30cm}
\includegraphics[width=3.5in]{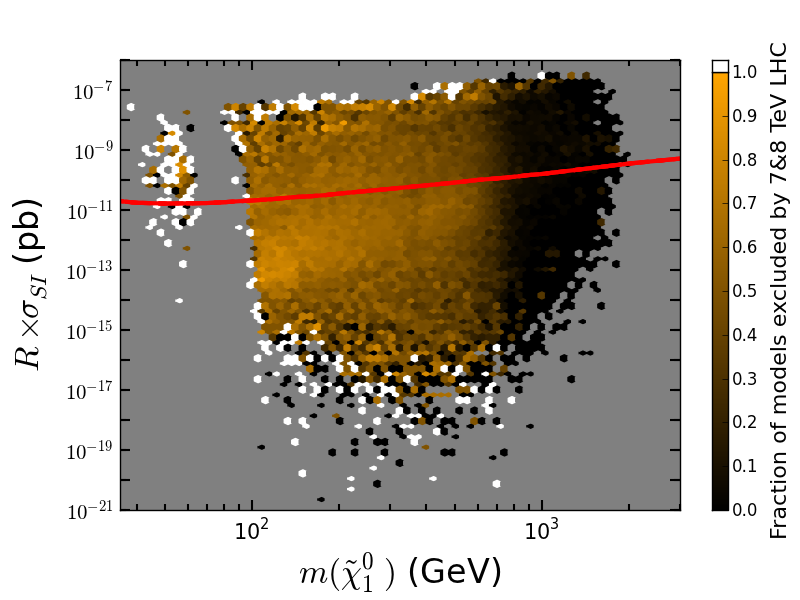}}
\vspace*{0.50cm}
\centerline{\includegraphics[width=3.5in]{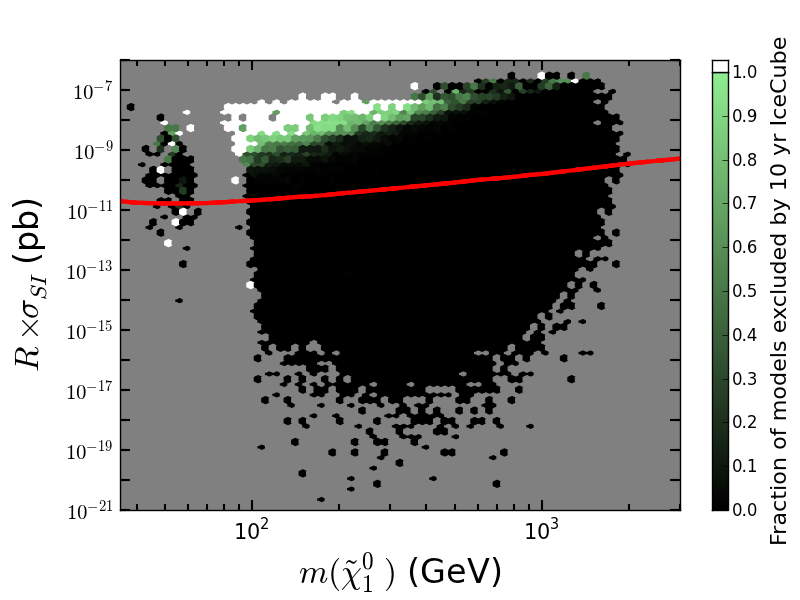}
\hspace{0.30cm}
\includegraphics[width=3.5in]{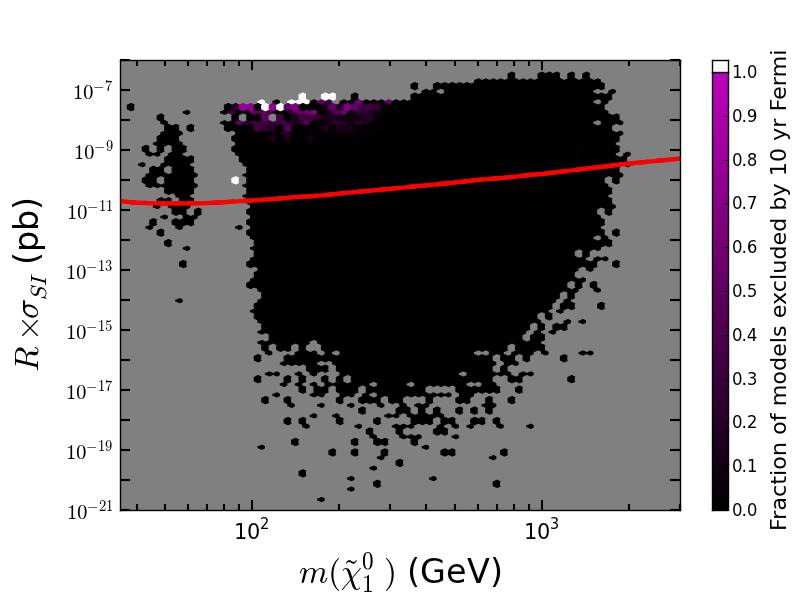}}
\vspace*{-0.10cm}
\caption{Comparisons of the search capabilities of various experiments projected into the LSP mass-scaled SI direct detection cross section plane, showing the fraction of models in each bin that can be excluded 
by CTA (top left), the LHC (top right), IceCube (bottom left) and \Fermi ~(bottom right). The expected SI cross-section upper limit from LZ with 1000 days of data is also depicted by the red curve.}
\label{figyy1}
\end{figure}

In order to study the complementarity between the various searches in more detail, it is instructive to project their individual capabilities onto parameter planes that are directly related to one of the search categories. This is particularly effective for visualizing how any given experiment's parameter space 
of interest is probed by other searches. As a first example of this, Fig.~\ref{figyy1} compares the 
search capabilities of various experiments in the familiar LSP mass-scaled SI direct detection cross section plane. Here we see the regions in this plane where individual experiments are most sensitive. In particular, we show the fraction of models excluded by CTA (top left), the LHC (top right), IceCube (bottom left) and \Fermi ~(bottom right) projected onto this plane. In each case, we also show the expected limit from the SI search at LZ. We see that both IceCube and \Fermi ~probe models with low LSP masses and large SI cross sections, where the LSP 
tends to be a bino-Higgsino admixture; this region is also accessible to the DD experiments such as LZ. On the other hand, CTA has access to the heavy LSP region, where there 
are a large fraction of relatively pure wino and Higgsino LSPs, while the LHC coverage is mostly concentrated (for now) on the relatively low mass LSP region. Interestingly, the 
LHC searches are not quite independent of the SI cross-section; a region of enhanced exclusion fraction is seen for SI cross sections near $\sim 10^{-13}$ pb. Models with SI cross 
sections in this region mostly have wino-like LSPs with light squarks, making them more likely to be observed by the LHC; wino-like LSPs with heavier squarks have a lower SI cross-section, while Higgsinos and mixed states tend to have a higher SI cross-section whether or not light squarks are present.

\begin{figure}[htbp]
\centerline{\includegraphics[width=3.5in]{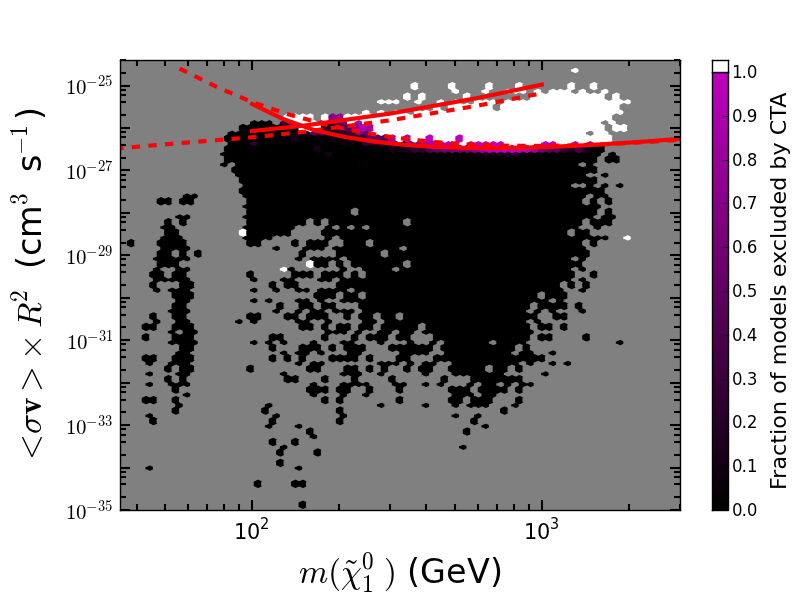}
\hspace{0.30cm}
\includegraphics[width=3.5in]{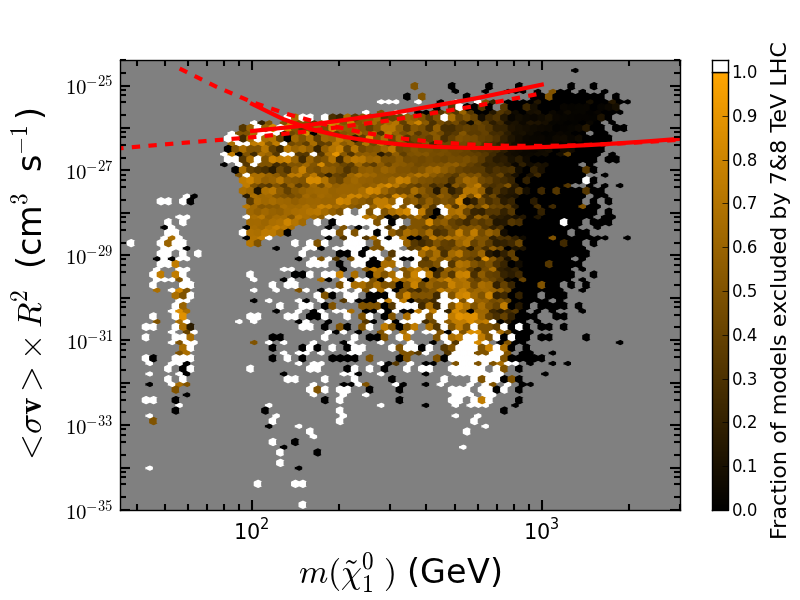}}
\vspace*{0.50cm}
\centerline{\includegraphics[width=3.5in]{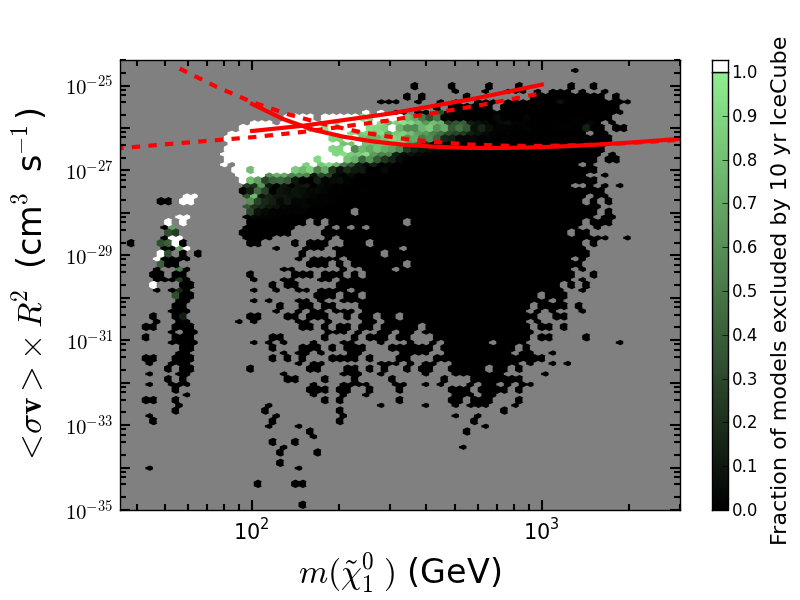}
\hspace{0.30cm}
\includegraphics[width=3.5in]{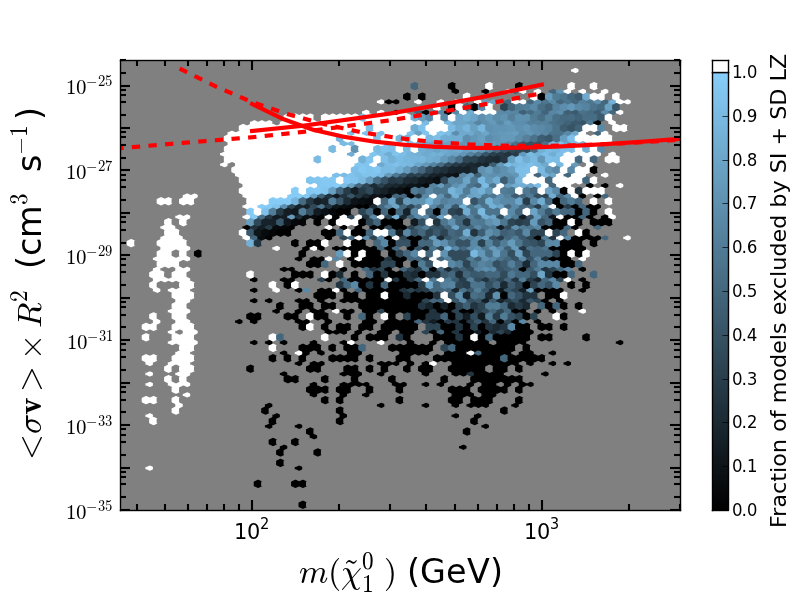}}
\vspace*{-0.10cm}
\caption{Comparisons of the search capabilities of various experiments in the LSP mass-LSP pair annihilation cross section plane, showing the fraction of 
models in each bin that can be excluded by CTA (top left), the LHC (top right), IceCube (bottom left) and LZ (bottom right). The expected bounds from \Fermi ~and 
CTA are also shown in each case as curves through the upper left and right parts of the figures, respectively. The dashed (solid) curves are 
for $100\%$ annihilation into $b\bar b~(W^+W^-)$ final states.}
\label{figyy2}
\end{figure}

We now project these results onto the plane which is most relevant for the DM ID searches. Figure~\ref{figyy2} again compares the search capabilities of various experiments, but now they are projected into the LSP mass-LSP pair annihilation cross section plane, in which the limits from \Fermi ~and CTA (with particular assumptions about the annihilation channels) can be presented directly. Here we see the fraction of models that can be excluded by searches at CTA (top left), the LHC (top right), IceCube (bottom left) and LZ (bottom right). The expected limits from \Fermi ~and CTA are also shown, represented by the curves penetrating the upper left- and right-hand side of the panels, respectively. Here the dashed (solid) curves correspond to the indirect detection limit obtained when the LSP pair annihilates exclusively into $b\bar b~(W^+W^-)$; we emphasize that a generic LSP in a pMSSM model may annihilate to many (often dozens of) different final states beyond these two simple 
cases. However, we do see that the generic pMSSM exclusion is well described by these limiting cases, as displayed for CTA in the upper left panel. Note again that CTA is primarily sensitive to models with LSP mass above 100 GeV and with \sigv relatively close to the thermal relic value.  As in the previous figure, the LHC 
is mainly effective in the lower LSP mass region. In addition, the LHC searches are seen to be particularly efficient along a thin line starting at annihilation cross sections of 
$\sim 10^{-29}\,\frac{cm^3}{s}$ for 100 GeV LSPs and increasing with LSP mass; this line corresponds to models with a pure wino LSP, which are subject to strong constraints from the 
LHC searches for heavy stable charged particles. (Note that this line is below the projected reach of the ID searches.) 
IceCube is seen to nicely complement CTA in the region with large cross sections but low LSP masses. The projection of the LZ coverage onto this plane is interesting with most of the 
exclusion appearing at lower LSP masses where the LSP pair annihilation cross section is also large. However, there is also an extended, somewhat diffuse region of substantial direct detection coverage 
throughout the entire right half of the parameter space as well as on the $Z,h$ funnel `island' at small LSP masses.


%
\begin{figure}[htbp]
\centerline{\includegraphics[width=3.5in]{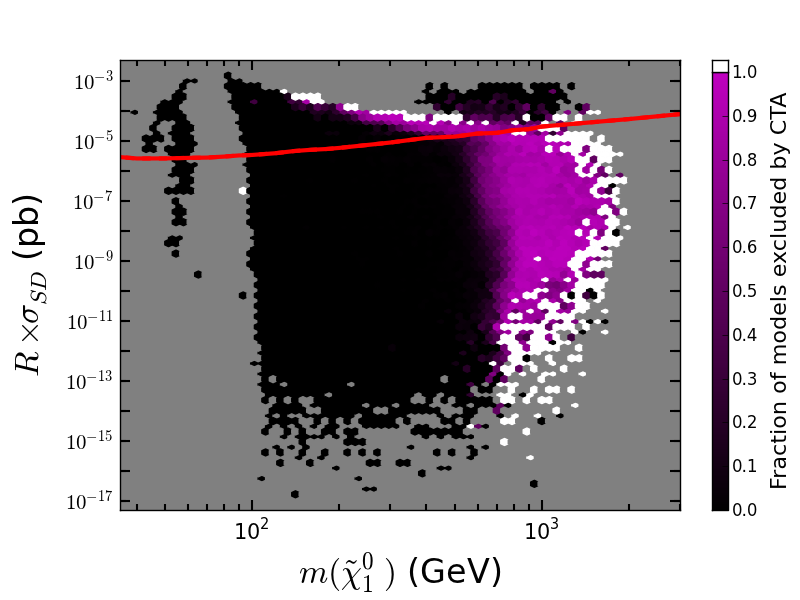}
\hspace{0.30cm}
\includegraphics[width=3.5in]{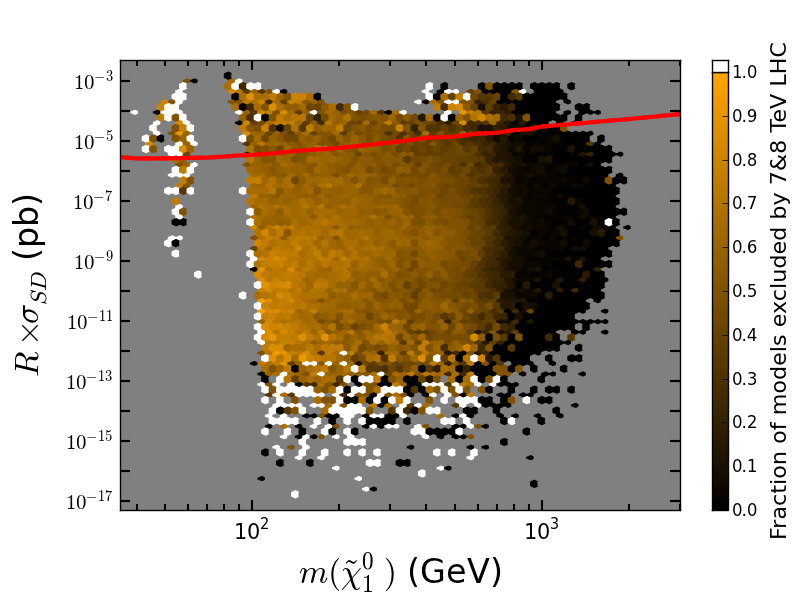}}
\vspace*{0.50cm}
\centerline{\includegraphics[width=3.5in]{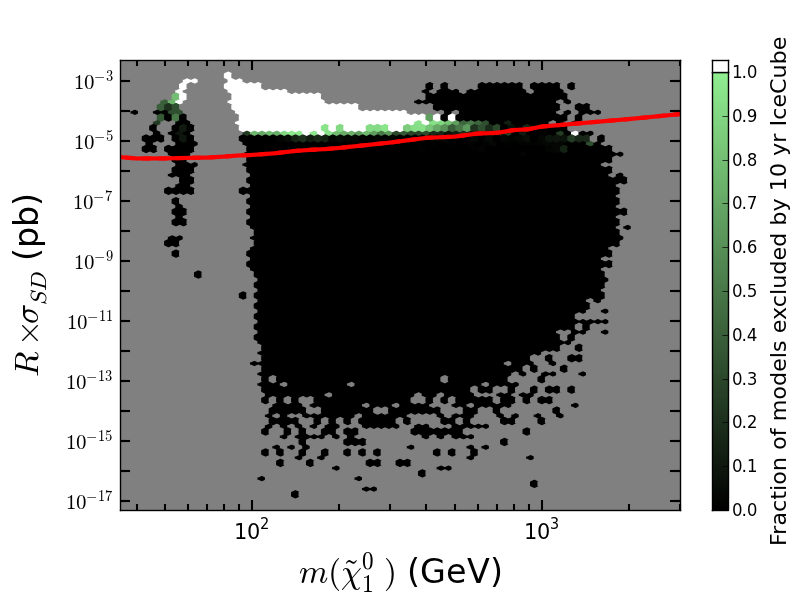}
\hspace{0.30cm}
\includegraphics[width=3.5in]{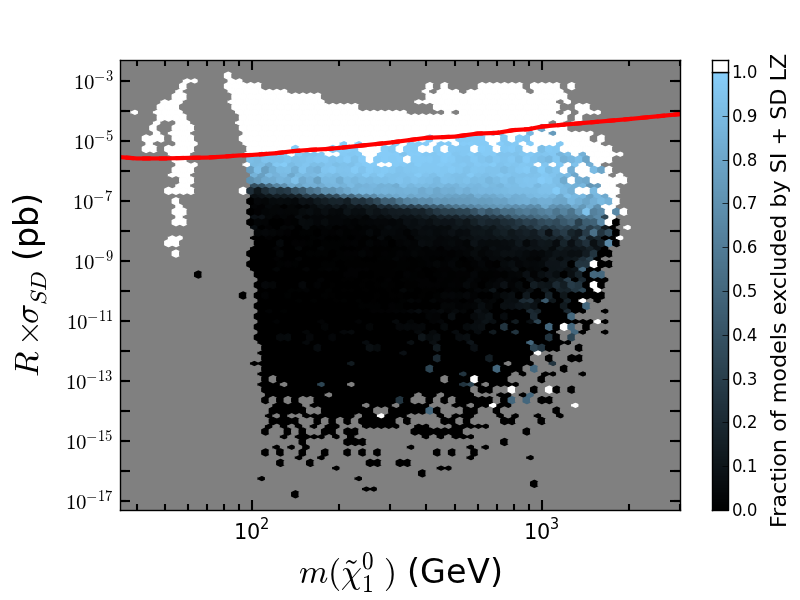}}
\vspace*{-0.10cm}
\caption{Comparisons of the search capabilities of various experiments in the LSP mass - proton SD cross section plane, showing the fraction of 
models in each bin that can be excluded by CTA (top left), the LHC (top right), IceCube (bottom left) and LZ (bottom right). The expected upper limit on the proton SD cross-section from LZ with 1000 days of data is represented by the red curve in each case.}
\label{figyy2x}
\end{figure}

Continuing along these lines, we also project our results onto the LSP mass - proton SD cross section plane, as shown in Fig.~\ref{figyy2x}. Spin-dependent direct detection experiments can place direct bounds in this plane (we show the expected SD constraint from LZ); indirect detection constraints from IceCube can also be represented in this plane, but depend on assumptions about annihilation channels and whether the LSP has come to equilibrium within the sun. Here we see that to first approximation the CTA and LHC sensitivities are essentially uncorrelated with the value of the SD cross section. IceCube is only sensitive to relatively light LSPs with large SD cross-sections, as would be expected from the IceCube bounds on specific annihilation channels. The lower right panel shows the fraction of models that can be excluded by LZ, including both the expected SD and SI results; the SD results are also described by the plotted red curve. We see that the SI LZ search is most sensitive to models with large spin-dependent cross sections, although the region of sensitivity extends far below the expected LZ SD sensitivity. In particular, we see that adding the LZ SI results removes all of the bino/Higgsino models in the $h/Z$-funnel region (LSP masses below $\sim$ 80 GeV).

\begin{figure}[htbp]
\centerline{\includegraphics[width=3.5in]{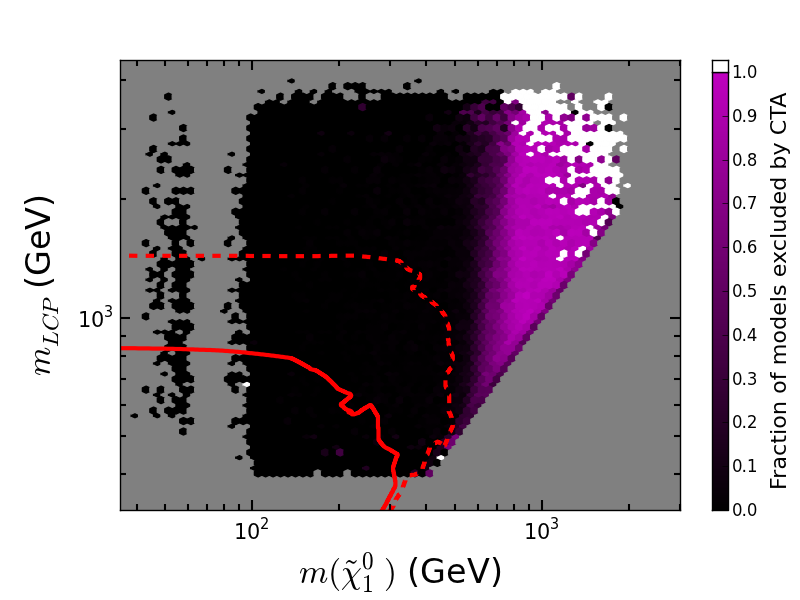}
\hspace{0.30cm}
\includegraphics[width=3.5in]{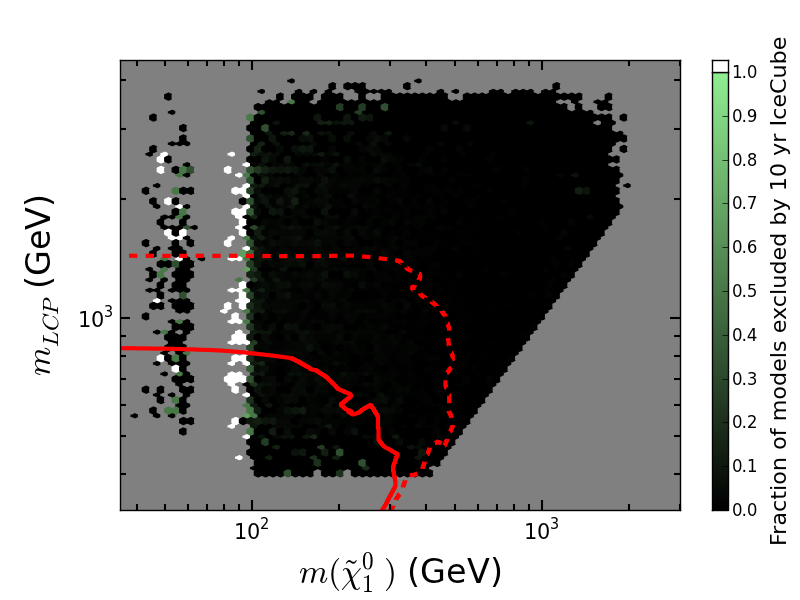}}
\vspace*{0.50cm}
\centerline{\includegraphics[width=3.5in]{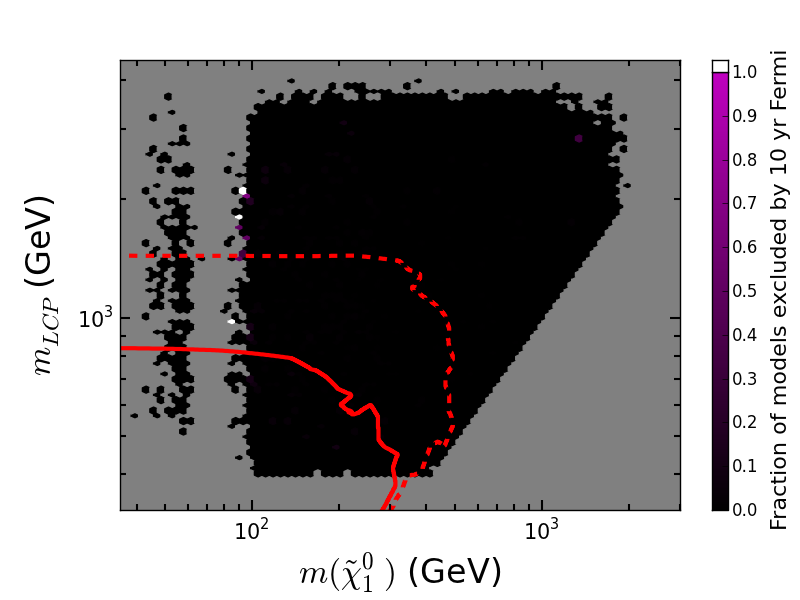}
\hspace{0.30cm}
\includegraphics[width=3.5in]{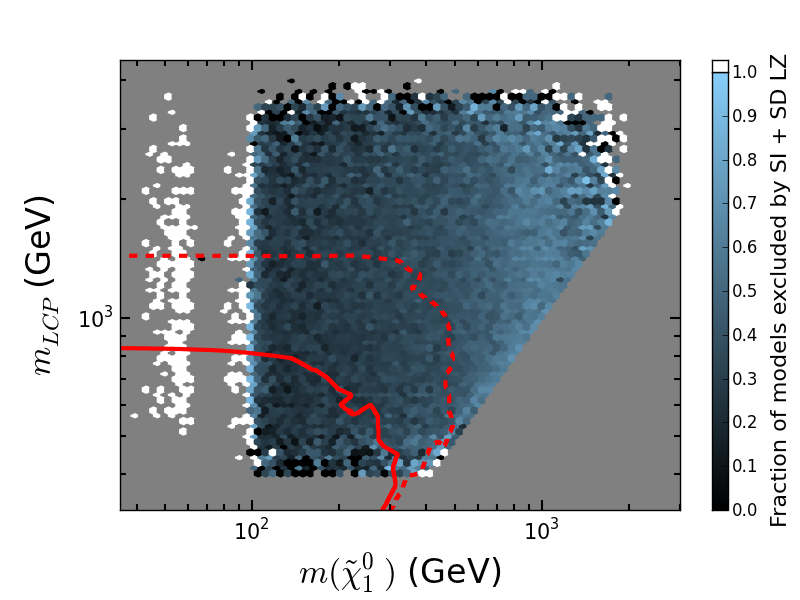}}
\vspace*{-0.10cm}
\caption{Comparisons of the search capabilities of various experiments in the LSP - Lightest (1st/2nd generation) Colored Sparticle (LCP) mass plane, showing the fraction of models in each bin that can be excluded by CTA (top left), IceCube
(top right), Fermi (bottom left) and LZ (bottom right). The dotted (solid) red curves show the ATLAS simplified model limits from the 20.3 fb$^{-1}$ jets plus MET analysis for 
gluinos (degenerate 1$^{st}$/2$^{nd}$ generation squarks).}
\label{figyy2y}
\end{figure}

Our last projection is displayed in Fig.~\ref{figyy2y}, which plots the LSP mass against the mass of the lightest colored sparticle (LCP){\footnote {Here the LCP is considered to be the lightest (1st or second generation) squark or gluino; the third generation squarks have somewhat different LHC signatures and are not included as the LCP.}}. The LHC SUSY 
searches, particularly the jets + MET channel, are directly applicable in this plane and are reasonably well described by simplified models for squarks and gluinos, as discussed above. Unsurprisingly, we see that the expected sensitivity for CTA, LZ, \Fermi, and IceCube depend strongly on the LSP mass, but are essentially independent of the LCP mass; the sensitivity of CTA appears to worsen when the LSP and LCP are nearly degenerate, most likely because the relic density in this case is reduced through co-annihilation rather than other mechanisms. On the other hand, the sensitivity of LZ is slightly enhanced in the compressed region, with the probable cause being that the presence of light squarks increases the scattering rate through squark exchange, compensating for the somewhat depleted relic density. 

\begin{figure}[htbp]
\centerline{\includegraphics[width=3.5in]{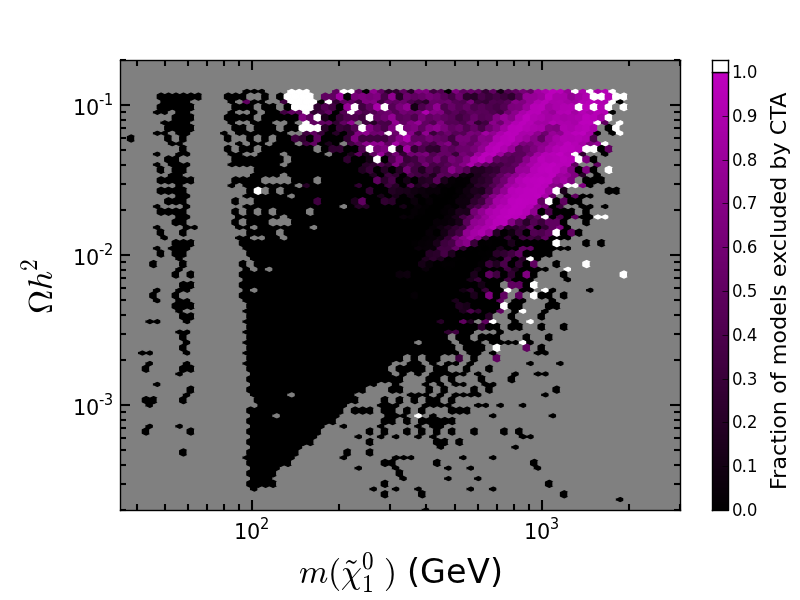}
\hspace{0.30cm}
\includegraphics[width=3.5in]{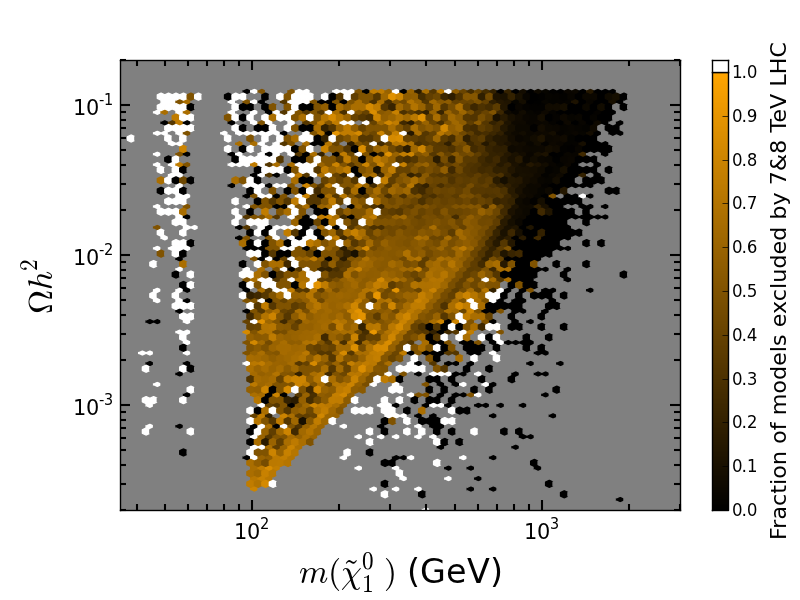}}
\vspace*{0.20cm}
\centerline{\includegraphics[width=3.5in]{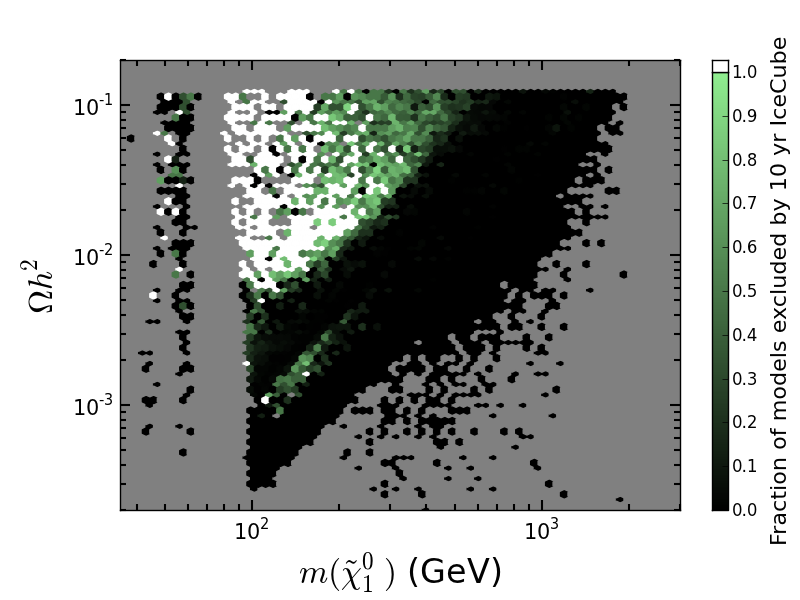}
\hspace{0.30cm}
\includegraphics[width=3.5in]{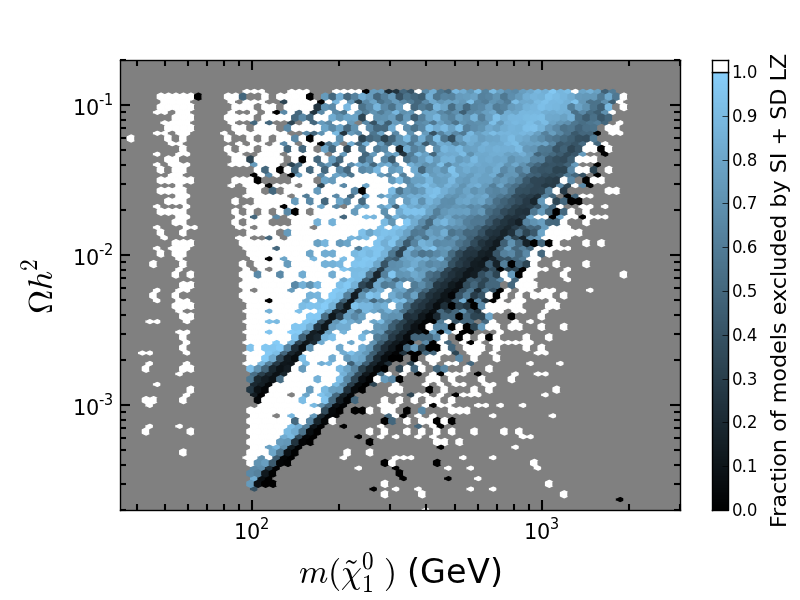}}
\vspace*{0.20cm}
\centerline{\includegraphics[width=3.5in]{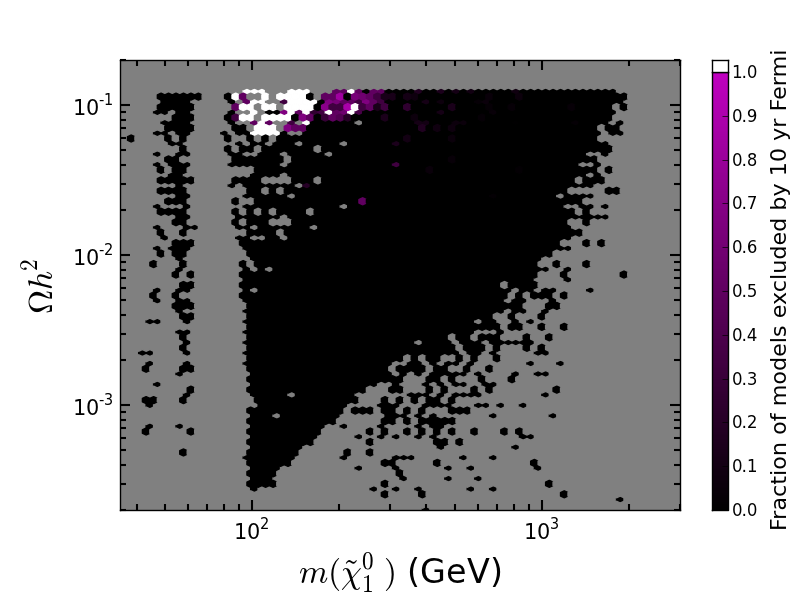}}
\vspace*{-0.10cm}
\caption{Comparisons of the search capabilities of various experiments in the LSP mass-relic density plane, showing the fraction of models that can be excluded by 
CTA (top left), the LHC (top right), IceCube (middle left), LZ (middle right) and \Fermi ~(bottom).}
\label{figyy3}
\end{figure}

Lastly, we now consider the thermal relic density-LSP mass plane. We note that with other factors constant, the sensitivity of direct detection and IceCube falls off linearly with decreasing relic density, while that of \Fermi~ and CTA falls off quadratically; of course the relic density plays no direct role in the LHC's sensitivity. In Fig.~\ref{figyy3} we display the fractions of models that can be excluded by CTA, the LHC, IceCube, LZ 
and \Fermi~ in this plane. Once again, we see that the regions constrained by the various experiments overlap significantly{\footnote {In the case of an actual 
DM {\it discovery}, the existence of a substantial overlap between the regions of experimental coverage within this parameter space will be very helpful when trying to determine 
the specific nature of the LSP.}}; for example, many different experiments will be sensitive to the ``well-tempered neutralino'' scenario. On the other hand, there are important cases where experiments complement each other to exclude a much larger fraction of models than could be seen by any one experiment. One example of this is the sensitivity of CTA to high-mass LSPs that are difficult to detect at the LHC; another is the sensitivity of the LHC to relatively light winos, which have a very low relic density and are generally missed by dedicated dark matter searches. Aside from being able to observe light winos through the production and decay of other sparticles, the LHC is also directly sensitive to them if their mass splitting is close to the pion mass, yielding a displaced decay that can appear in searches for disappearing tracks or heavy stable charged particles. We also note several other features apparent in this figure: First, CTA, as expected, has excellent sensitivity to most of the models with LSP masses above $\sim 250$ GeV that saturate the relic density. However, for larger masses, the CTA coverage also extends to relic densities as much as a factor of $\sim 10$ or more below the WMAP/Planck value. \Fermi~ is seen to cover only the low LSP mass region with a relic density not far from the thermal value, while the IceCube sensitivity extends to much lower relic density provided the LSP mass 
is below $\sim 500$ GeV or so. LZ has sensitivity throughout this plane but does best for LSP masses below $\sim 300$ GeV, even for models with very low relic density. Of course, even for LSP masses up to 1-2 TeV, the LZ sensitivity remains reasonably good. As noted already, the LHC is presently seen to be effective mainly for LSP masses below $\sim 500-600$ GeV. The LHC coverage is relatively uniform with respect to the relic density, but of course the fraction of models excluded is very high in the case of very light LSPs.  
Of course, we again remind the reader that extending the LHC energy to 14 TeV will substantially improve its sensitivity to heavy LSPs, as we will see below.

\begin{figure}[htbp]
\centerline{\includegraphics[width=7.0in]{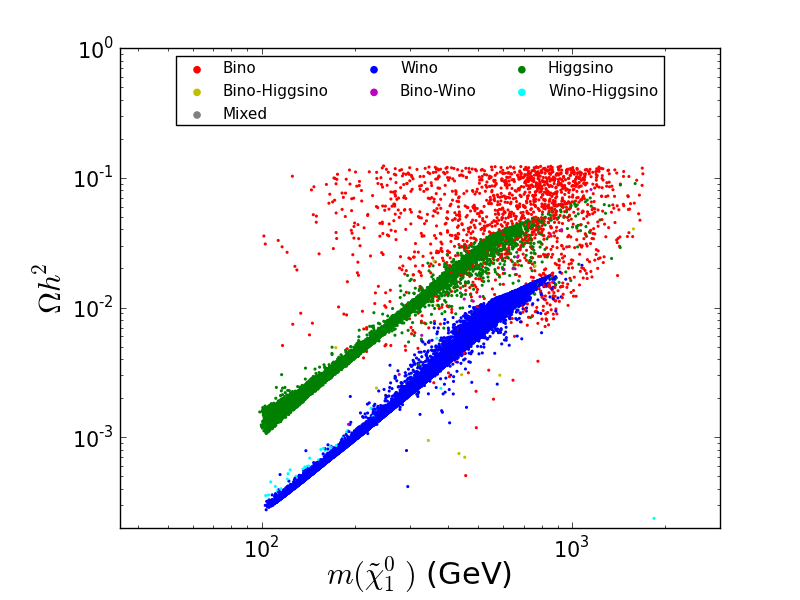}}
\vspace*{-0.10cm}
\caption{Thermal relic density as a function of the LSP mass for the pMSSM models, surviving after the expected constraints from all the searches are taken into account, color-coded by the 
electroweak properties of the LSP. Compare with Fig.~\ref{fig00}. }
\label{figzz}
\end{figure}

Finally, Fig.~\ref{figzz} displays the impact of combining all the expectations for the different searches in the $\Omega h^2$-LSP mass plane; this should be compared with Figure~\ref{fig00}, showing the original 
model set before the search sensitivities have been applied. Here we see that ($i$) the models that were in the light $h$ and $Z$-funnel regions have completely evaporated through 
a combination of the SI and SD LZ analyses, ($ii$) the well-tempered neutralinos are now seen to have completely disappeared, mostly due to LZ and IceCube, with additional help from \Fermi, ($iii$) the possibility of almost pure 
Higgsino or wino LSPs even approximately saturating the relic density has vanished due to CTA, ($iv)$ the mixed wino-Higgsino models, due to a combination of measurements, 
have also completely disappeared, ($v$) the only models remaining which {\it do} saturate the WMAP/Planck relic density are those with binos with $A$-scalar resonant annihilation or co-annihilation. 
($vi$) We find that $\sim 75.5\%$ of the pMSSM model sample will have been excluded (or observed) by at least one of the searches considered in this paper.

\section{Complementarity with the 14 TeV LHC}

\begin{figure}[htbp]
\centerline{\includegraphics[width=3.5in]{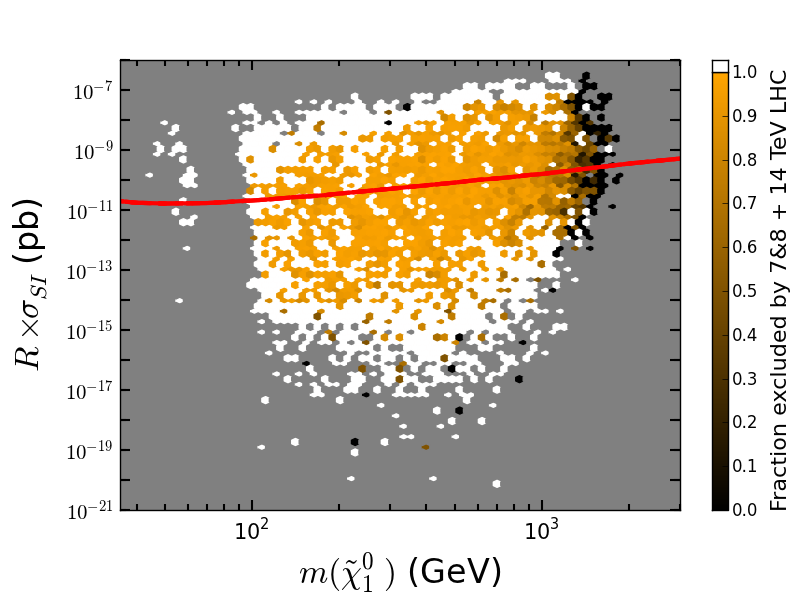}
\hspace{0.30cm}
\includegraphics[width=3.5in]{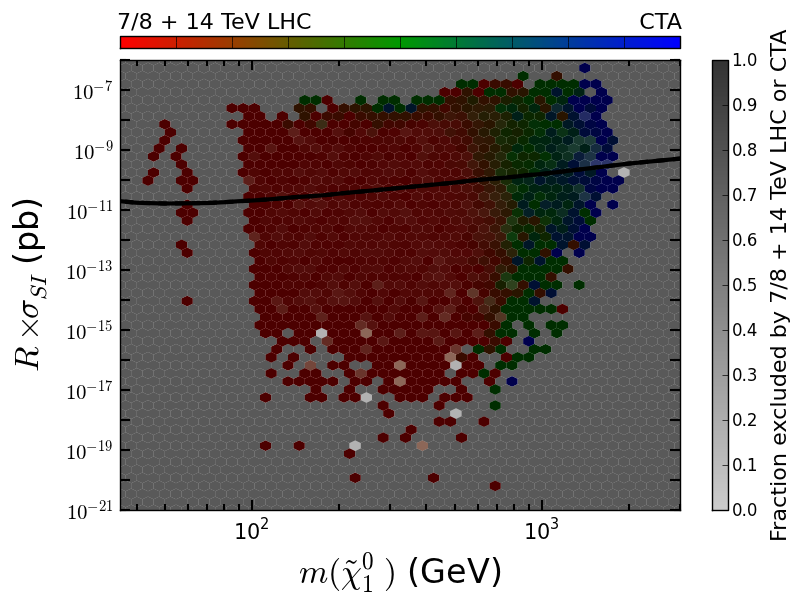}}
\vspace*{-0.10cm}
\caption{The fraction of models with the correct Higgs mass which are excluded by the combination of the 14 TeV jets + MET and the 0$\ell$ + 1$\ell$ stop searches with 300 fb$^{-1}$, shown in 
the LSP mass-scaled SI cross section plane (left panel) and a comparison between the fractions of models excluded by the LHC and CTA in this plane (right panel). The expected limit on the Xenon SI cross section from LZ is also shown in both cases.}
\label{14TeVComp}
\end{figure}

We now consider the effect of adding 14 TeV jets + MET and corresponding 0-lepton and 1-lepton stop searches with 300 fb$^{-1}$ of integrated luminosity to the full set of 7+8 TeV searches considered previously. Here, we restrict our analysis to the subset of $\sim$ 45k models with $m_h = 126\pm 3$ GeV, due to computational limitations. We remind the reader that both the LHC and dedicated dark matter results are essentially independent of the Higgs mass, so that the results for this subset should effectively reproduce those we would get for the full model set, albeit with lower statistics. The left panel of Fig.~\ref{14TeVComp} shows the reach of the combined LHC 
searches in the LSP mass - SI cross-section plane for models with $m_h = 126\pm 3$ GeV. Comparing this figure to the upper right panel of Fig.~\ref{figyy1}, we see two key changes with 
the inclusion of the expected 14 TeV searches. First, in both cases the effectiveness of the LHC searches is seen to fall off sharply above a particular LSP mass, since above this limit the spectrum 
is generically either too heavy or too compressed to be observed. Adding the 14 TeV searches effectively doubles this cutoff, from $\sim$ 700 GeV to $\sim$ 1400 GeV, so that most LSPs in 
our model set can be excluded given colored sparticle masses below $\sim$ 2-3 TeV. Second, we see that the LHC now has sensitivity to a very high fraction (but not all) of the models with LSPs 
lighter than this cutoff. The large fraction of models which can be excluded by the 14 TeV data is unsurprising, since our chosen scan range for the sparticle masses was 
designed to ensure that most models would be (at least kinematically) accessible at the 14 TeV LHC.

The right panel of Fig.~\ref{14TeVComp} shows a comparison between the reaches of the LHC and CTA in this plane, analogous to the lower right panel of Figure~\ref{figxx}. We see that 
the LHC and CTA sensitivities now exhibit a sizable region of overlap. However, the blue region on the far right edge of the panel shows that CTA will be sensitive to LSP masses 
beyond the reach of the LHC. We also note that LSPs heavy enough to be seen by CTA are generally too heavy to be detected in direct (e.g. monojet) searches, so that the LHC is sensitive to 
these models by observing other (mostly colored) sparticles. It is therefore important to note that CTA has the potential to exclude winos and Higgsinos with a nearly thermal relic density  \textit{regardless} of the characteristics of the rest of the sparticle spectrum. Of course, the improved reach of the LHC at 14 TeV means that there is also an increasing overlap between the LHC and 
LZ, however we see that (as before) the LHC searches are mostly independent of the SI cross-section. (The white areas at the edges of the panel generally result from low statistics in 
those regions, increasing the likelihood that all of the models in a given bin will be excluded). Of course, from a discovery perspective, the increased overlap means that there is more potential for a signal to be observed by two, or even all three, experiments, 
which would greatly aid in characterizing the LSP and other model properties. Finally, we note that not only has the total fraction of models that can be excluded by the combined experiments increased 
dramatically (from 75.5\% to 98.0\%) as a result of the 14 TeV LHC reach, but the fraction of models not seen by the LHC which are covered by direct or indirect detection has increased slightly (from 54.8\% to 59.3\%), since more of the undetected models have heavy LSPs and are therefore likely to be covered by CTA.

\section{Conclusion}

The results presented here directly lead us to a number of interesting conclusions that are already apparent and that we expect to strengthen in the future as more data is collected 
at the LHC:

\begin{itemize}

\item{Even if the LSP {\it does not} make up all of the DM, it can still be observed in both direct and indirect detection experiments, as well as neutrino experiments such as IceCube. 
Of course, searches at the LHC are not influenced by the LSP relic density.}

\item{The set of models remaining after all the searches are performed that saturate the thermal relic density consist almost uniquely of those with (co)annihilating bino LSPs.}

\item{SI direct detection, CTA, and the LHC do most of the heavy lifting in terms of complementary searches covering the pMSSM parameter space.}

\item{Multiple/overlapping searches allow for extensive parameter space coverage that will be of particular importance if a DM signal is observed.}

\item{Most of the experiments are seen to provide complementary probes of the pMSSM parameter space.} 

\item{The strength of the LHC component in these searches increases significantly with the inclusion of the 14 TeV LHC sensitivity. However, \eg, the jets + MET search is dependent on the 
rest of the model spectrum, and therefore does not provide complete coverage of any given LSP scenario, in contrast to dedicated DM searches which rely more directly on the LSP properties.}

\end{itemize}

In summary, the pMSSM provides an excellent tool for studying complementarity between different approaches to the search for dark matter. Hopefully, DM will soon be discovered so that 
we can employ the complementary probes discussed above to ascertain its nature.

\section*{Acknowledgments}

The authors would like to thank A. Barr, J. Buckley, D. Cote, J. Feng, K. Matchev, G. Redlinger, and T. Tait for useful discussions. This work was supported by the Department 
of Energy, Contracts DE-AC02-76SF00515, DE-AC02-06CH11357, and DE-FG02-12ER41811, the Department of Energy Office of Science Graduate Fellowship Program (DOE SCGF), made possible in part by the American Recovery and 
Reinvestment Act of 2009, administered by ORISE-ORAU under contract no. DE-AC05-06OR23100, and the National Science Foundation under grant PHY-0970173.

\end{document}